\documentclass[twocolumn,amsmath,amssymb,aps,pra,floatfix]{revtex4}
\usepackage{bm}
\usepackage{amsfonts}
\usepackage{amsmath}
\usepackage{graphicx}
\usepackage{float}

\begin{document} 
\title{Manipulating neutral particles in Bessel beams: from rings, through fixed helices to 3D traps}
\author{Tomasz Rado\.zycki}
\email{t.radozycki@uksw.edu.pl}
\affiliation{Faculty of Mathematics and Natural Sciences, College of Sciences, Institute of Physics, Cardinal Stefan Wyszy\'nski University, W\'oycickiego 1/3, 01-938 Warsaw, Poland} 
\begin{abstract}
The motion of neutral, polarizable atoms (also called neutral particles in this work) in the field of the Bessel beam is considered. It is shown in the numerical way, that the Bessel rings, i.e., the regions of high energy concentration can trap particles of positive polarizability (atoms in red-detuned beams). This trapping occurs only in the plane perpendicular to the wave propagation, and the motion along the beam is unrestricted. When the beam is superposed with the plane wave of the same frequency propagating in the same direction, the particles are guided along helices, fixed in space. The shape of these helices depends on the parameters characterizing the electromagnetic fields but not on the initial state of guided particles. Depending on the vorticity of the Bessel beam, these helices can be made left or right-handed. In the special case of zero vorticity, the helices get degenerated to the true, three-dimensional rings, which can serve as $3D$ traps. The emerging structure of potential valleys can be applied to parallel guidance or capture several independent atoms, each in its own trap.
\end{abstract}
\maketitle

\section{Introduction}\label{int}

\begin{figure}[h]
\begin{center}
\includegraphics[width=0.38\textwidth,angle=0]{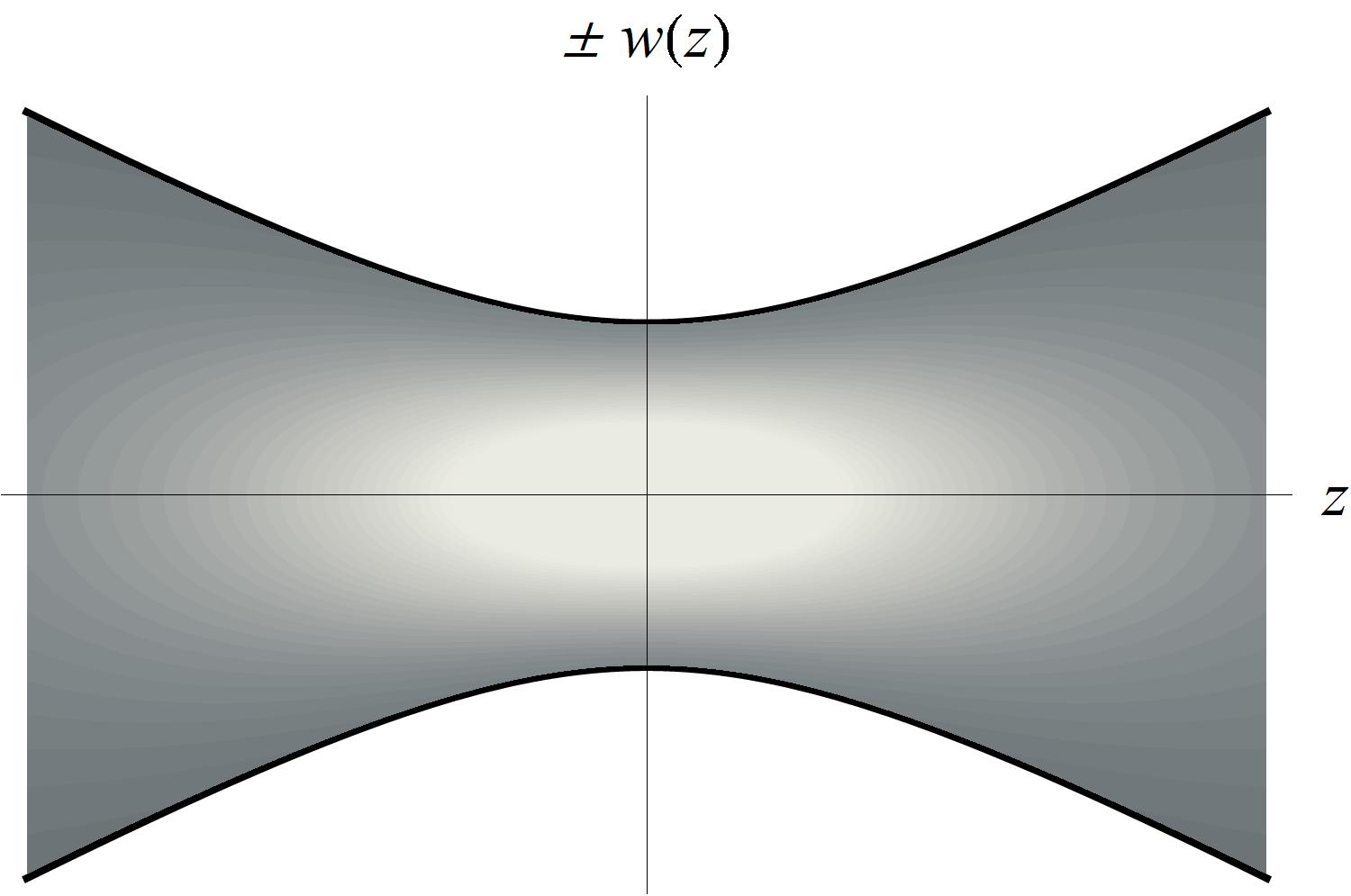}
\end{center}
\caption{The light intensity of a typical Gaussian beam. Bright region corresponds to the higher concentration of energy. The beam propagates along the $z$-axis, and $w(z)$ stands for the spot size.}
\label{gauss}
\end{figure}

The idea of manipulating small neutral objects with the use of light beams was practically born around 1970 with an experimental work by Ashkin~\cite{ashkin1} in which the author demonstrated the possibility of controlling the motion of particles by the beam of radiation. In his experiment, transparent dielectric micrometer-sized spheres placed in the water were given acceleration by means of a Gaussian laser beam. In such a beam, the law of light refraction leads to the appearance of the gradient force dragging a given object into the area of higher or lower electromagnetic energy density. If the refractive index of a sphere with respect to the surrounding medium is greater than unity, it is pulled into the area of higher energy density. If, however, the material of the sphere is optically rarer, it is repelled from such area (for further details and references see the review articles~\cite{neu,dho1,die}). The light intensity distribution of a typical Gaussian beam is presented in Fig.~\ref{gauss}. It explicitly shows the white region in the center which represents the increased energy density attracting the high-index spheres. 

Apart from this property the acceleration in the direction of beam propagation due to radiation pressure exerted on the sphere pushes it along the beam. In order to avoid this effect and construct a real $3D$ trap, one can apply for instance two opposite laser beams~\cite{ashkin1}. Instead of the second beam of radiation the gravitational force can also be used to stabilize the dielectric spheres or for instance liquid drops leading to the levitation phenomenon~\cite{ashkin2,ashkin3,ashkin4}.

Similar mechanism based on the presence of gradient forces has been applied to neutral atoms in beams red-detuned from the characteristic atomic lines. The appearance of the optical force is based on the Stark effect~\cite{dk} or equivalently on polarizability of atoms. The radiation pressure on such atoms is strongly reduced and the force pushing them along the beam becomes smaller than the gradient force pulling atoms backward into the bright region of Fig.~\ref{gauss}. The practical implementation of this type of a trap called the optical tweezer has been reported in~\cite{chu} and later in~\cite{miller}. Such traps are also realized in the intersecting beams (see for instance~\cite{adams}). A similar blue-detuned trap was also shown to work~\cite{dav,sheng,frie}. In the case of far-detuned beams the atomic excitations can be avoided and the dipole force becomes dominating~\cite{adams1, odtna}. Optical tweezers have found many applications in distant areas of activity, from physics to biology or medicine  (for instance~\cite{ste,fazal,pad,woe,bowpa,grier1}). Good theoretical background is given in~\cite{brad}.

The Gaussian beam, as shown in Fig.~\ref{gauss}, exhibits relatively simple structure with one energy concentration zone localized at beam's waist. Much richer possibilities of trapping or guiding particles are provided by beams of the so-called structured light~\cite{sgg,andrews,lee}, where the word `structure' mainly refers to phase. One can enumerate here for instance  Laguerre-Gaussian~\cite{lg,lg2}, Bessel~\cite{durnin1,durnin2,vg},  Mathieu~\cite{ma} or Airy~\cite{siv} beams. An up-to-date summary of the results on `twisted light' is gathered in the review work~\cite{babik}. 
In the present paper we concentrate on Bessel beams, which belong to the class of non-diffracting beams~\cite{bou,yu,nd}, and on their ability to manipulate neutral polarizable particles~\cite{arlta}. Pure Bessel beams would carry the infinite energy, so in practical applications one naturally deals with Bessel beams truncated in the radial direction, which however is irrelevant for the present work. 

For another type of beams of similar kind, namely Bessel-Gaussian beams, the trapping effects have been revealed by showing the existence of a waveguiding potential, used consequently to investigate the Bose-Einstein condensate~\cite{arltb}  

The mechanism of trapping neutral particles as atoms is similar to that spoken of above in the context of Gaussian beam, but the spatial structure of the Bessel beam opened a variety of ways to manipulate microparticles~\cite{arltc,vgc,asd,ta}. 

Since it would be difficult to talk about the capture of relativistic particles in traps of the potential depth measured in electronovolts (which will be estimated later), it is quite sufficient to consider in this paper the non-relativistic motion. 

Assuming that the atomic dipole moment is proportional to the external electric field, i.e.
\begin{equation}
{\bm d}=\alpha {\bm E},
\label{de}
\end{equation}
where $\alpha$ typically depends on the driving frequency, one gets the equation of motion in the form
\begin{equation}
m\ddot{\bm r}= ({\bm d}\cdot {\bm \nabla}){\bm E}=\frac{1}{2}\,\alpha {\bm \nabla}(E^2).
\label{rr1}
\end{equation}
The appearance of the optical force is then caused by the non-homogeneity of electric field. As stems from the theory of the Stark effect, for the red-detuned beam the polarizability $\alpha$ is positive, and for the blue-detuned negative~\cite{odtna}. Is is interesting to mention that the similar potential with negative value of $\alpha$ is obtained for charged particles in the ponderomotive potential and, therefore, there are some similarities between the motion of neutral atoms and charged particles as for instance electrons in the Bessel beam~\cite{ibb1,ibb2}.

In the following section the potential for the polarizable particle in a pure Bessel beam is analyzed and the equation of motion~(\ref{rr1}) is examined. The trajectories of both red-detuned and blue-detuned atoms are obtained numerically in Sec.~\ref{tratab}. This section serves as a kind of an introduction for more complicated system of fields. The analysis of the motion of atoms placed in the combination of a Bessel beam and a plane wave is dealt with in Sec.~\ref{bbpl}. Up to our knowledge this configuration of fields has not been dealt with before as a structure for trapping and guiding neutral particles. 

Assuming that these two waves are precisely tuned, it is shown that if the Bessel beam has nonzero vorticity the conservative potential emerging after taking the average over rapid oscillations, has minima in the form of helices which are fixed in space (Sec.~\ref{pote}). These helices can serve to transport atoms independently on their initial velocity as shown numerically in Sec.~\ref{hm}. Depending on the topological vortex number, several atoms can be transported in parallel, independently of each other. The twist of the helices and their pitch are connected with the value of this number (higher vorticity produces looser helix). In the special case of zero vorticity  it is shown in Sec.~\ref{trl} that the considered configuration of waves leads to the appearance of a system of real $3D$ toroidal traps, equally spaced which form a kind of a lattice of traps.

\section{Atoms in the Bessel beam}\label{bb}

The electric field in the monochromatic Bessel beam of frequency $\omega$ propagating along the $z$-axis has the form:
\begin{subequations}\label{elebb}
\begin{align}
E_x&=E_0\big[\kappa_-\cos(k_z z-\omega t+(M+1)\varphi)J_{M+1}(k_\perp \rho)\nonumber\\
&+\kappa_+\cos(k_z z-\omega t+(M-1)\varphi)J_{M-1}(k_\perp \rho)\big], \label{elebbx}\\
E_y&=E_0\big[\kappa_-\sin(k_z z-\omega t+(M+1)\varphi)J_{M+1}(k_\perp \rho)\nonumber\\
&-\kappa_+\sin(k_z z-\omega t+(M-1)\varphi)J_{M-1}(k_\perp \rho)\big], \label{elebby}\\
E_z&=2 E_0\sin(k_z z-\omega t+M\varphi)J_M(k_\perp \rho). \label{elebbz}
\end{align}
\end{subequations}
For non-relativistic motion, we are dealing with, the influence of the magnetic field may be neglected. The following notation has been used above:
\begin{equation}
\kappa_\pm=\frac{\sqrt{k_z^2+k_\perp^2}\pm k_z}{k_\perp}, \label{notkappa}
\end{equation}
where $k_z^2+k_\perp^2=\omega^2/c^2$, $\rho=\sqrt{x^2+y^2}$ stands for the radial variable and $\varphi$ for the azimuthal angle. Each particular beam is characterized by an integer number $M$ which represents the vorticity of the field (strictly speaking the vortex topological number equals $M-1$). The quantities $J_M$ refer to the Bessel functions of the first kind.

\subsection{Potential}\label{potb}

As mentioned in the Introduction for the motion of polarizable particles in the beam described with the equation~(\ref{rr1}) the role of the potential is played by the quantity
\begin{eqnarray}
V(&&\!\!\!\!\!\!\!\rho,\varphi,z,t)=-\frac{1}{2}\,\alpha {\bm E}(\rho,\varphi,z,t)^2\label{vpot}\\
&&\!\!=-\frac{1}{2}\,\alpha E_0^2 \big[\kappa_-^2J_{M+1}(k_\perp\rho)^2+\kappa_+^2J_{M-1}(k_\perp\rho)^2\nonumber\\
&&\;\;\;+2J_{M+1}(k_\perp\rho)J_{M-1}(k_\perp\rho)\cos(2(k_z z-\omega t +M\varphi))\nonumber\\
&&\;\;\;+4J_M(k_\perp\rho)^2\sin^2(k_z z-\omega t+M\varphi)\big],
\nonumber
\end{eqnarray}
where the identity $\kappa_+\kappa_-=1$ has been utilized. Upon introducing the following dimensionless variables and auxiliary constants:
\begin{subequations}\label{diml}
\begin{align}
&\xi_x=k_\perp x,\;\;\;\xi_y=k_\perp y,\;\;\;\xi=k_\perp \rho,\;\;\;\zeta=k_z z,\;\;\;\; \tau=\omega t, \label{dimlvar}\\
&\sigma_\perp=\frac{c k_\perp}{\omega},\;\;\; \sigma_z=\frac{c k_z}{\omega},\;\;\; \beta=\frac{\alpha E_0^2\sigma_\perp^2}{mc^2},\;\;\; \gamma=\frac{\alpha E_0^2\sigma_z^2}{mc^2}. \label{dimlcon}
\end{align}
\end{subequations}
the equation of motions can be rewritten in the suitable form~(\ref{rrdiml}). The quantities $\beta$ and $\gamma= (\sigma_z/\sigma_\perp)^2\beta$ which satisfy the identity
\begin{equation}
\beta^2+\gamma^2=\left(\frac{\alpha E_0^2}{mc^2}\right)^2,
\label{bg}
\end{equation}
are typically very small since they refer the electric energy to the rest energy of a given particle in motion. For a generic Bessel beam one has $\sigma_\perp\ll\sigma_z$ which leads to the condition $\beta\ll\gamma$. The constants $\kappa_\pm$ now read
\begin{equation}
\kappa_\pm=\frac{1\pm\sigma_z}{\sigma_\perp}.
\label{kasi}
\end{equation}
It should be noted that for a beam propagating along the $z$-axis in the positive direction (as in the present case) $\kappa_+\gg 1$ and $\kappa_-\ll 1$. If the beam run in the opposite direction, these two quantities would switch roles.

Denoting
\begin{eqnarray}
{\cal V}({\bm \xi},\zeta,\tau)=&&\!\!\!\!\!\!-\kappa_-^2J^2_{M+1}(\xi)-\kappa_+^2J^2_{M-1}(\xi)\nonumber\\
&&\!\!\!\!\!\!\!-2J_{M+1}(\xi)J_{M-1}(\xi)\cos(2(\zeta+\tau+M\varphi))\nonumber\\
&&\!\!\!\!\!\!\!-4J^2_M(\xi)\sin^2(\zeta+\tau+M\varphi)
\label{vcal}
\end{eqnarray}
where ${\bm \xi}=[\xi_x,\xi_y]$ (${\cal V}({\bm \xi},\zeta,\tau)$ should be alternatively understood as ${\cal V}(\xi,\varphi,\zeta,\tau)$ if needed), the equations of motion can be given the form
\begin{subequations}\label{rrdiml}
\begin{align}
\ddot{\xi}_x(\tau)&=-\frac{\beta}{2}\,\nabla_{\xi_x}{\cal V}({\bm \xi},\zeta,\tau),\label{rrdimlx}\\
\ddot{\xi}_y(\tau)&=-\frac{\beta}{2}\,\nabla_{\xi_y}{\cal V}({\bm \xi},\zeta,\tau),\label{rrdimly}\\
\ddot{\zeta}_x(\tau)&=-\frac{\gamma}{2}\,\nabla_\zeta{\cal V}({\bm \xi},\zeta,\tau).\label{rrdimlz}
\end{align}
\end{subequations}
In what follows, ${\cal V}$ is called the `potential'. The true potential, obviously, has the additional and very small  factor $\beta/2$ (or $\gamma/2$ if one accounts for different scaling of the $\zeta$-axis).

Since the electric field is time dependent the total energy is not conserved but rather satisfies
\begin{equation}
\frac{d}{d\tau}\,{\cal E}=\frac{d}{d\tau}\left[\frac{\dot{\xi}_x^2}{2\beta}+\frac{\dot{\xi}_y^2}{2\beta}+\frac{\dot{\zeta}^2}{2\gamma}+\frac{1}{2}\, {\cal V}\right]=\frac{1}{2}\,\frac{\partial}{\partial\tau} {\cal V}.
\label{cm1}
\end{equation}
On the other hand in the considered setup there is a constant of motion related to the invariance with respect to the symmetry which consists in the simultaneous shifting of $\zeta$ and $\varphi$ ($\zeta\mapsto \zeta +M\varepsilon$, $\varphi\mapsto \varphi-\varepsilon$):
\begin{equation}
{\cal C}=\frac{1}{\beta}\left(\xi_x\dot{\xi}_y-\dot{\xi}_x\xi_y\right)+\frac{M}{\gamma}\, \dot{\zeta}.
\label{const}
\end{equation}
In the particular case of vanishing $M$ this constant reduces to the third component of the angular momentum (AM) of the particle. This happens because such a beam does not carry AM: the spin and orbital momenta are opposite. Consequently, the potential looses its $\varphi$ dependence and no force in the azimuthal direction is exerted on the particle. For $M\neq 0$ such a force does exist and the transfer of the AM from the field to the particles is observed. As will be seen below, the non-conservation of the AM in this case happens in short-time scale: AM understood as a mean, where fast oscillations are eliminated, is conserved.

Potential (\ref{vcal}) depends on time but in typical situations the oscillations of the the electromagnetic field are very rapid as compared to the eventual oscillatory motion of the atom. It is too heavy to keep up with the evolution of the electric field for optical frequencies or even many orders of magnitude lower. For instance for small oscillations the harmonic force can be written as
\begin{equation}
{\bm F}=\frac{\alpha}{2}\, ({\bm x}\cdot {\bm \nabla}){\bm \nabla} E^2
\label{fham}
\end{equation}
which leads to atomic frequencies satisfying the condition
\begin{equation}
\frac{\omega_\mathrm{at}^2}{\omega^2}\sim \frac{\alpha E_0^2}{2mc^2} (k_\perp^2, k_\perp k_z, k_z^2)=\frac{1}{2} (\beta,\sqrt{\beta\gamma}, \gamma)\ll 1.
\label{wat}
\end{equation}
This enables to calculate the average the `potential' over short-time oscillations of the electric field and consequently to set $\cos(\ldots)\mapsto 0$ and $\sin^2(\ldots )\mapsto 1/2$, simplifying the expression for ${\cal V}$:
\begin{eqnarray}
{\cal V}_a({\bm \xi},\zeta)=&&\!\!\!\!\!\!\langle {\cal V}({\bm \xi},\zeta,\tau)\rangle_\tau=\label{vcals}\\
&&\!\!\!\!\!\!\!-\kappa_-^2J^2_{M+1}(\xi)-\kappa_+^2J^2_{M-1}(\xi)-2J^2_M(\xi),\nonumber
\end{eqnarray}
$a$ standing for `averaged'. This approximation is tested below also in the numerical way. The obtained potential is no longer $\varphi$ dependent, which means that particle's AM is conserved, if understood as an average over rapid oscillations. ${\cal V}_a$ can be further simplified if one remembers that $\kappa_+\gg 1$ and $\kappa_-\ll 1$. From among the above three terms the second one is dominant.

\subsection{Trajectories of atoms}\label{tratab}

\begin{figure}[h]
\begin{center}
\includegraphics[width=0.475\textwidth,angle=0]{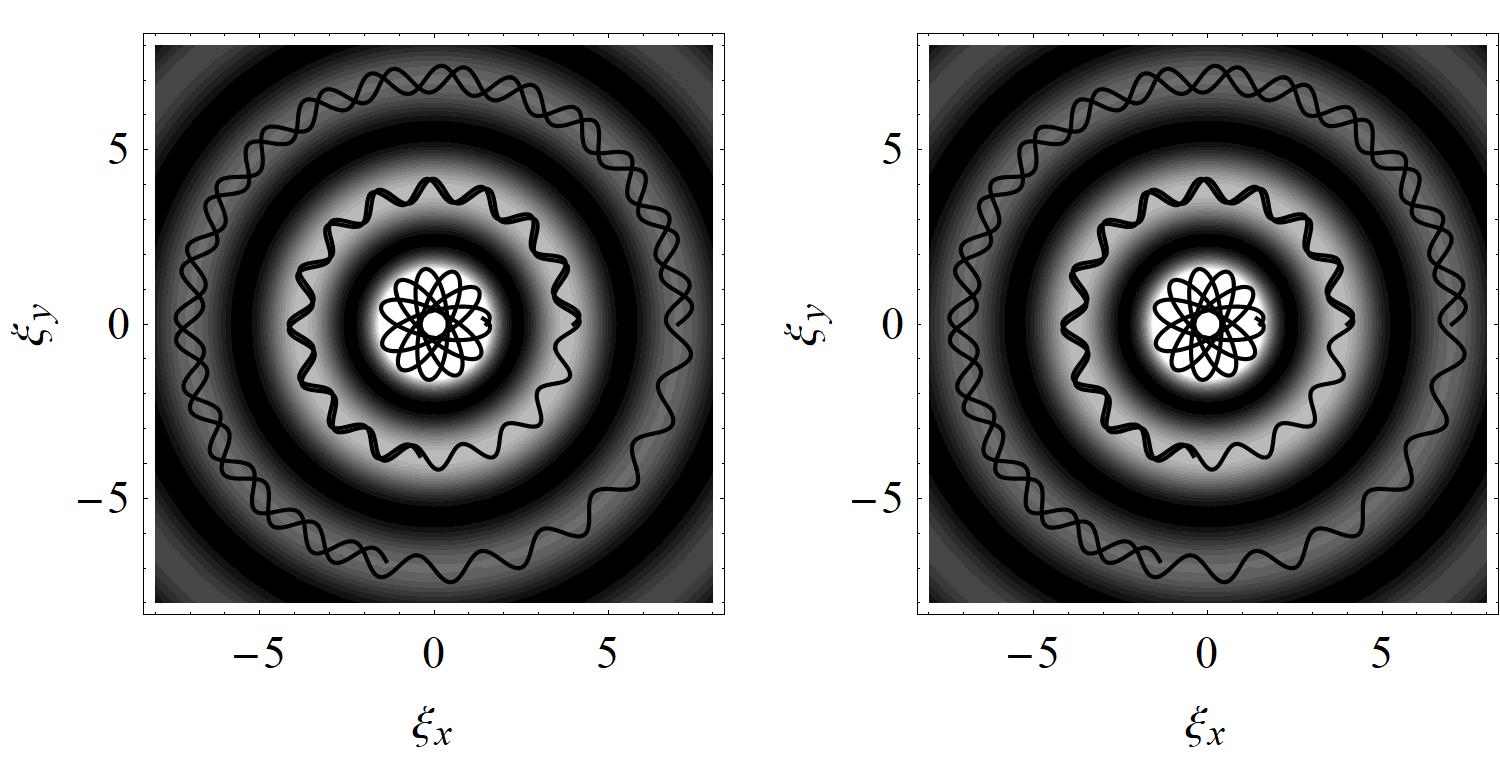}
\end{center}
\caption{The exemplary trajectories of neutral atoms in the Bessel beam with $M=1$ obtained from the approximate potential ${\cal V}_a$ (left diagram) and full potential ${\cal V}$ (right diagram) in the red-detuned case. In both cases $\beta=5\times 10^{-4}$, $\sigma_\perp=0.1$. }
\label{comp-black}
\end{figure}

The motion described with equations~(\ref{rrdiml}) even determined by the simplified potential ${\cal V}_a$ cannot be solved analytically. The plots of the numerical solutions are presented in Fig.~\ref{comp-black}. The left diagram shows the motion of atoms in the $(\xi_x, \xi_y)$ plane in the Bessel beam for $M=1$ according to the equations~(\ref{rrdimlx}) and~(\ref{rrdimly}) with the approximate potential ${\cal V}_a$. On the right the trajectories with the same initial conditions but with the use of the full potential ${\cal V}$ are depicted. For $M\neq 1$ the beam has nonzero vorticity and consequently is devoid of the central core, but the motion of atoms in the Bessel rings is identical.

Two conclusions can be drawn from these graphs. First, the trajectories are practically identical, which emphasizes the validity of the approximation used. Second, the particles have tendency to stay in the regions of high energy concentration (bright rings). This phenomenon is well known from trapping neutral atoms by light beams as spoken of in the Introduction and is connected with the sign of the polarizability $\alpha$. 

For $\alpha>0$, which corresponds to red-detuned wave, the effect of trapping particles within Bessel rings is obtained as seen in Fig.~\ref{comp-black}. For tiny oscillations, the radial motion is harmonic with the small frequency of $\beta^{1/2}\omega$. In the realistic situation of, say, ${}^7Li$ atoms in the laser beam of intensity $I=10^{11}\,\mathrm{W/cm^2}$ the potential depth of the subsequent rings (still for $M=1$) can be estimated to $14.4\,\mathrm{eV}$, $2.4\,{\mathrm eV}$, $1.3\,\mathrm{eV}$, and so on, where the value of the atomic polarizability was roughly taken as $\alpha=10^3$ in atomic units ($1\,\mathrm{a.u.}\approx 1.65\times 10^{-41}\, \mathrm{C^2m^2/J}$), where calculations show that $164<\alpha<1991$ dependent on the wave frequency~\cite{dinampol}. Since the potential is proportional to the electric field squared, for lower laser intensities the depths are proportionally reduced. Hence, the potential wells may be thought of as sufficiently deep, and the trap should operate effectively. For other values of $\alpha$ the obtained depths are accordingly modified.

\begin{figure}[h]
\begin{center}
\includegraphics[width=0.475\textwidth,angle=0]{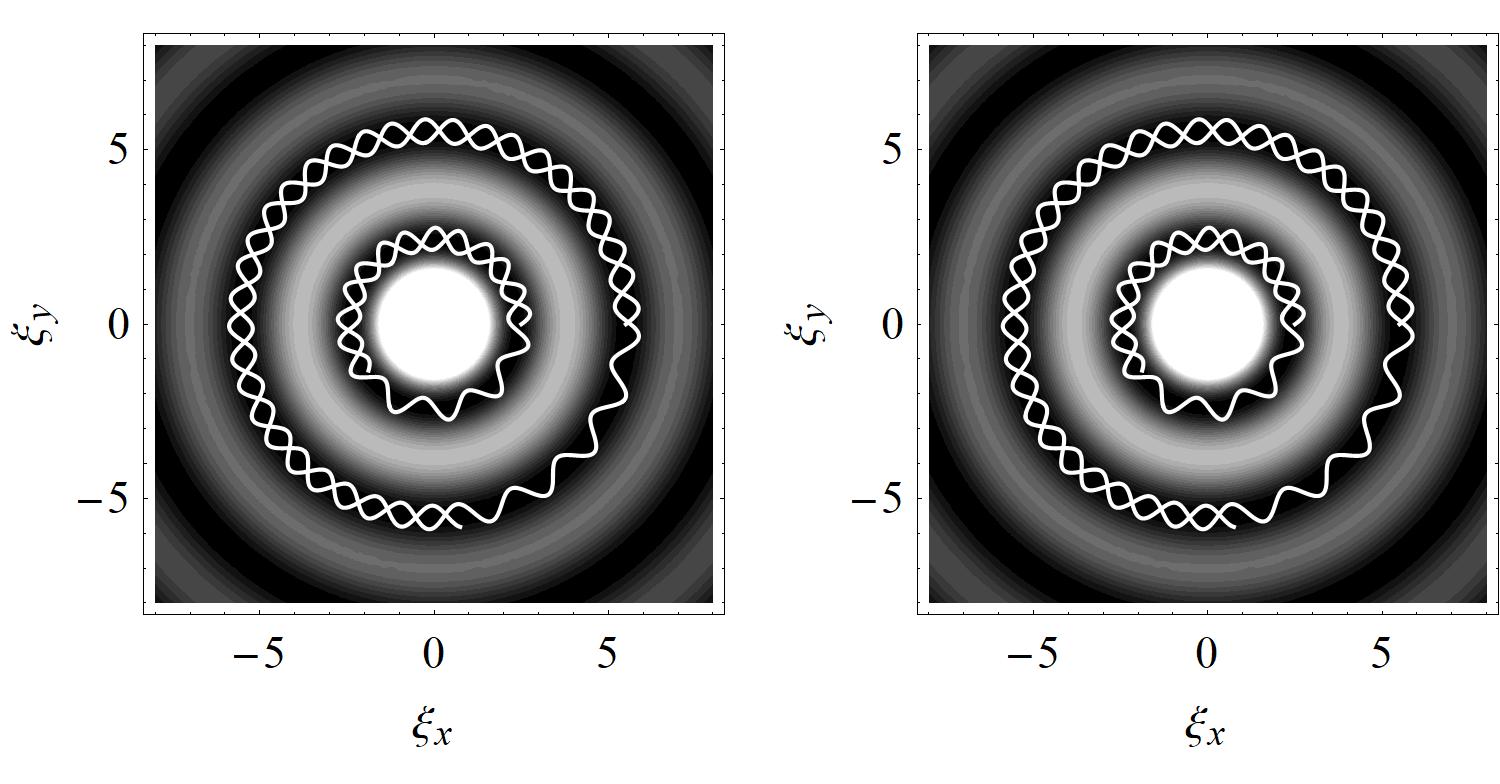}
\end{center}
\caption{Same as in Fig.~\ref{comp-black} but for blue-detuning ($\beta=-5\times 10^{-4}$).}
\label{comp-white}
\end{figure}

For the blue-detuned wave the polarizability reverses its sign, so the atoms should gather in the space between Bessel rings (dark regions). That it is actually the case may be seen from numerical calculations presented in Fig.~\ref{comp-white}. As mentioned in the Introduction, this situation applies as well (in the qualitative sense) to charged particles in the ponderomotive potential created by the Bessel beam~\cite{ibb1}. Similar trajectories were also obtained in~\cite{ibb2} without referring to the ponderomotive potential. However, the trapping mechanism for charged particles considered in these two cited works is entirely different from that considered in the present work in the context of neutral atoms. For the case of microparticles the resembling accumulation in the Bessel rings was observed experimentally in~\cite{asd, brown}.

Larger objects (for instance of the size of micrometers) are subject to Brownian motion, which can result in hopping between various rings as seen in experiments~\cite{brown}. On the other hand for small particles the tunneling effect can come into play when solving the problem in quantum mechanics. However, this phenomenon seems to be negligible due to the thickness of the barrier between subsequent rings. In any case, both effects can be minimized either by dilution of the air or cooling down particles, which makes the barrier even thicker and higher and, naturally, by manipulating the intensity of the waves.

\begin{figure}[h]
\begin{center}
\includegraphics[width=0.285\textwidth,angle=0]{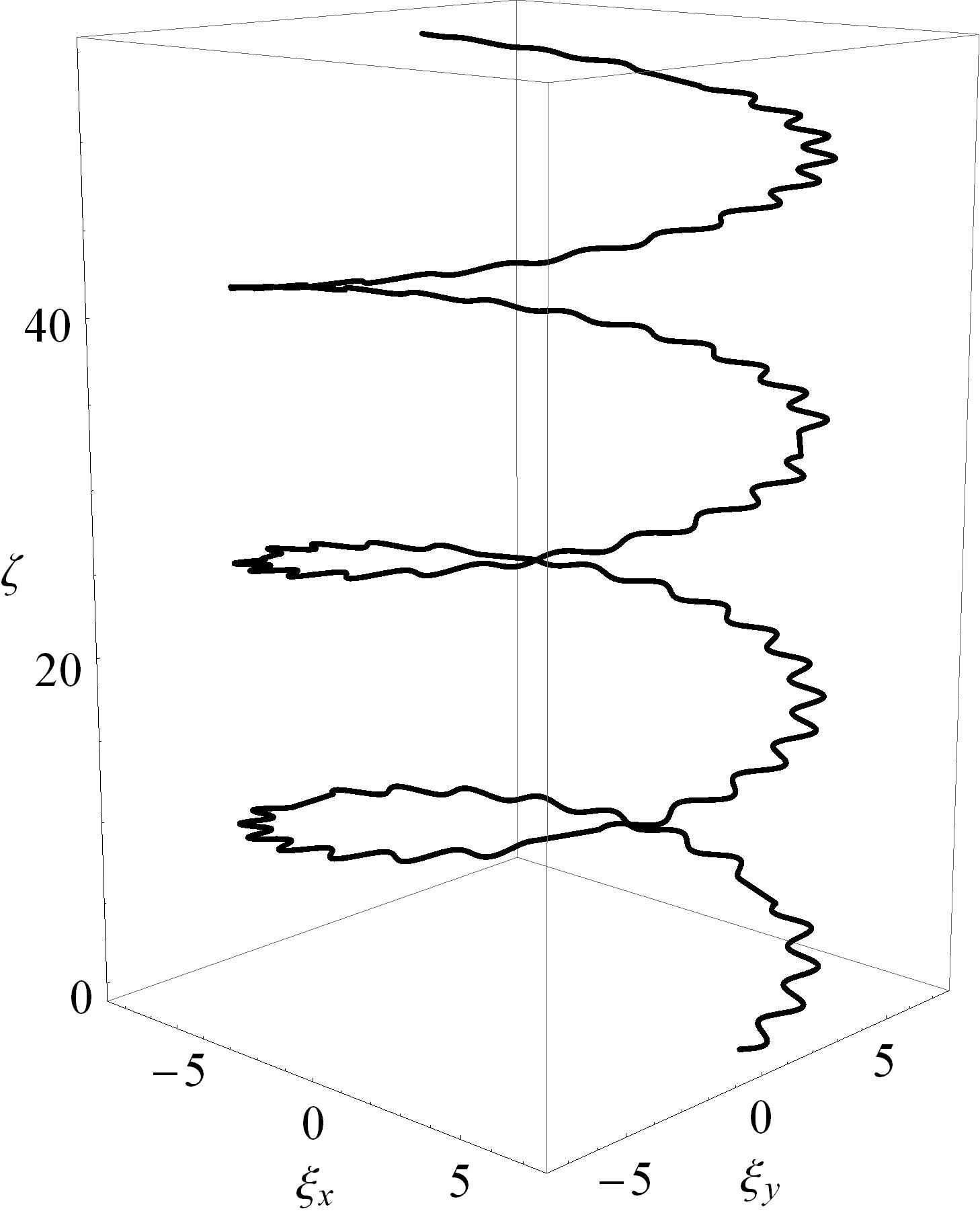}
\end{center}
\caption{A typical $3D$ trajectory of a neutral polarizable particle moving in the Bessel beam obtained by solving equations~(\ref{rrdiml}). The values of the parameters are the same as in Fig.~\ref{comp-black}.}
\label{3db}
\end{figure}

The motion in the $\zeta$-direction is subject to small (and fast) oscillations, which, however, can be smoothed out if the above-discussed approximation has been applied. This is reflected by the fact, that due to the time averaging, potential~(\ref{vcal}) is no longer $\zeta$ dependent. Therefore, in the considered case the motion in the direction parallel to the principal axis is uniform. Therefore, the motion whose projection on the $(\xi_x,\xi_y)$ plane is shown in Figures~\ref{comp-black} and~\ref{comp-white} in reality runs along a helix, which, however, does not have a fixed character as it depends on the particle initial state. A typical $3D$ trajectory obtained from the numerical calculations is depicted in Fig.~\ref{3db}. In the next section it will be shown that the superposition of a Bessel beam with a plane wave leads to the appearance of {\em fixed} helices whose dimensions in space are unequivocally determined and not affected by atom's initial conditions.

\section{Atoms in the Bessel beam and plane wave field}\label{bbpl}

\subsection{Potential}\label{pote}

Interesting potential structures arise if a Bessel beam is subject to the superposition with a plane wave of the same  $\omega$ propagating along the same axis. For charged particles moving in the ponderomotive potential generated by the similar arrangement of waves they were obtained in~\cite{ibb1}. As we shall see, choosing both frequencies to be identical allows to get rid of the time dependence and get the nontrivial conservative potential ${\cal V}_a$. In the case of the small detuning, one can produce the slowly varying potential (the helices to be spoken of below start to rotate about the $\zeta$-axis), which can turn out to be useful for manipulating particles, but remains outside the scope of interest of this work.

Let us assume the electric field of the plane wave in the form:
\begin{equation}
{\bm E}=qE_0({\bm e}_x \cos(k z-\omega t+\psi)-\chi {\bm e}_y \sin(k z-\omega t+\psi)),
\label{elepw}
\end{equation}
with $q$ standing for the measure of the relative strength of both fields and the choice of $\chi$ fixes the polarization (the value $\chi=1$ corresponds to positive helicity, $\chi=-1$ to negative and $\chi=0$ to linear polarization along the $x$-axis). According to~(\ref{rr1}), the potential governing the motion of polarizable particles can be written as
\begin{widetext}
\begin{eqnarray}
{\cal V}({\bm \xi},\zeta,\tau)=&&\!\!\!\!\!\!\!-\kappa_-^2J^2_{M+1}(\xi)-\kappa_+^2J^2_{M-1}(\xi)-2J_{M+1}(\xi)J_{M-1}(\xi)\cos(2(\zeta+\tau+M\varphi))-4J^2_M(\xi)\sin^2(\zeta+\tau+M\varphi)\nonumber\\
&&\!\!\!\!\!\!\!-q\kappa_-J_{M+1}(\xi)\left[(1+\chi)\cos\left(\frac{1+\sigma_z}{\sigma_z}\zeta-2\tau+(M+1)\varphi+\psi\right)+(1-\chi)\cos\left(\frac{1-\sigma_z}{\sigma_z}\zeta-(M+1)\varphi+\psi\right)\right]\nonumber\\
&&\!\!\!\!\!\!\!-q\kappa_+J_{M-1}(\xi)\left[(1-\chi)\cos\left(\frac{1+\sigma_z}{\sigma_z}\zeta-2\tau+(M-1)\varphi+\psi\right)+(1+\chi)\cos\left(\frac{1-\sigma_z}{\sigma_z}\zeta-(M-1)\varphi+\psi\right)\right]\nonumber\\
&&\!\!\!\!\!\!\!-q^2\left[1+(\chi^2-1)\sin^2\left(\frac{1}{\sigma_z}\,\zeta-\tau+\psi\right)\right],
\label{vcalt}
\end{eqnarray}
\end{widetext}
where again the dimensionless quantities~(\ref{diml}) have been introduced. After time averaging in order to get rid of the rapid oscillations the potential simplifies to
\begin{eqnarray}
&&\!\!\!\!\!\!\!{\cal V}_a({\bm \xi},\zeta)=\langle {\cal V}({\bm \xi},\zeta,\tau)\rangle_\tau\label{vcalts}\\
&&\!\!\!\!\!\!=-\kappa_-^2J^2_{M+1}(\xi)-\kappa_+^2J^2_{M-1}(\xi)-2J^2_M(\xi)-\frac{1+\chi^2}{2}\,q^2\nonumber\\
&&\!\!\!\!-q(1-\chi)\kappa_-J_{M+1}(\xi)\cos\left(\frac{1-\sigma_z}{\sigma_z}\,\zeta-(M+1)\varphi+\psi\right)\nonumber\\
&&\!\!\!\!-q(1+\chi)\kappa_+J_{M-1}(\xi)\cos\left(\frac{1-\sigma_z}{\sigma_z}\,\zeta-(M-1)\varphi+\psi\right),
\nonumber
\end{eqnarray}
where $q^2(1+\chi^2)/2$ is an inessential constant and may be omitted. Since $\kappa_+\gg 1\gg\kappa_-$, the further simplification of this expression is possible as in the previous section, leading to
\begin{eqnarray}
&&\!\!\!\!\!\!\!{\cal V}_{as}({\bm \xi},\zeta)=-\kappa_+^2J^2_{M-1}(\xi)\label{vcaltsa}\\
&&\!\!\!\!-q(1+\chi)\kappa_+J_{M-1}(\xi)\cos\left(\frac{1-\sigma_z}{\sigma_z}\,\zeta-(M-1)\varphi+\psi\right),
\nonumber
\end{eqnarray}
This formula holds both for $\chi=1$ and $\chi=0$. It is then seen that in this approximation the (positive) circular polarization and the linear polarization will lead to the identical motion of the particle and $\chi=1$ merely entails the modification of the parameter $q$ ($q\mapsto 2q$), which eventually allows to appropriately reduce the plane wave intensity. Therefore, in our numerical analysis the value $\chi=0$ is simply set and to account for the circular polarization one simply divides $q$ by $2$. This rule is exact for the potential(\ref{vcaltsa}) and almost exact in the case of~(\ref{vcalts}).

The simplified form is better suited for a qualitative analysis, but the numerical calculations in Sec.~\ref{hm} are performed with the use of the full form~(\ref{vcalt}). As it was in the case of the pure Bessel beam, they are in perfect harmony with each other.

\begin{figure}[h]
\begin{center}
\includegraphics[width=0.4275\textwidth,angle=0]{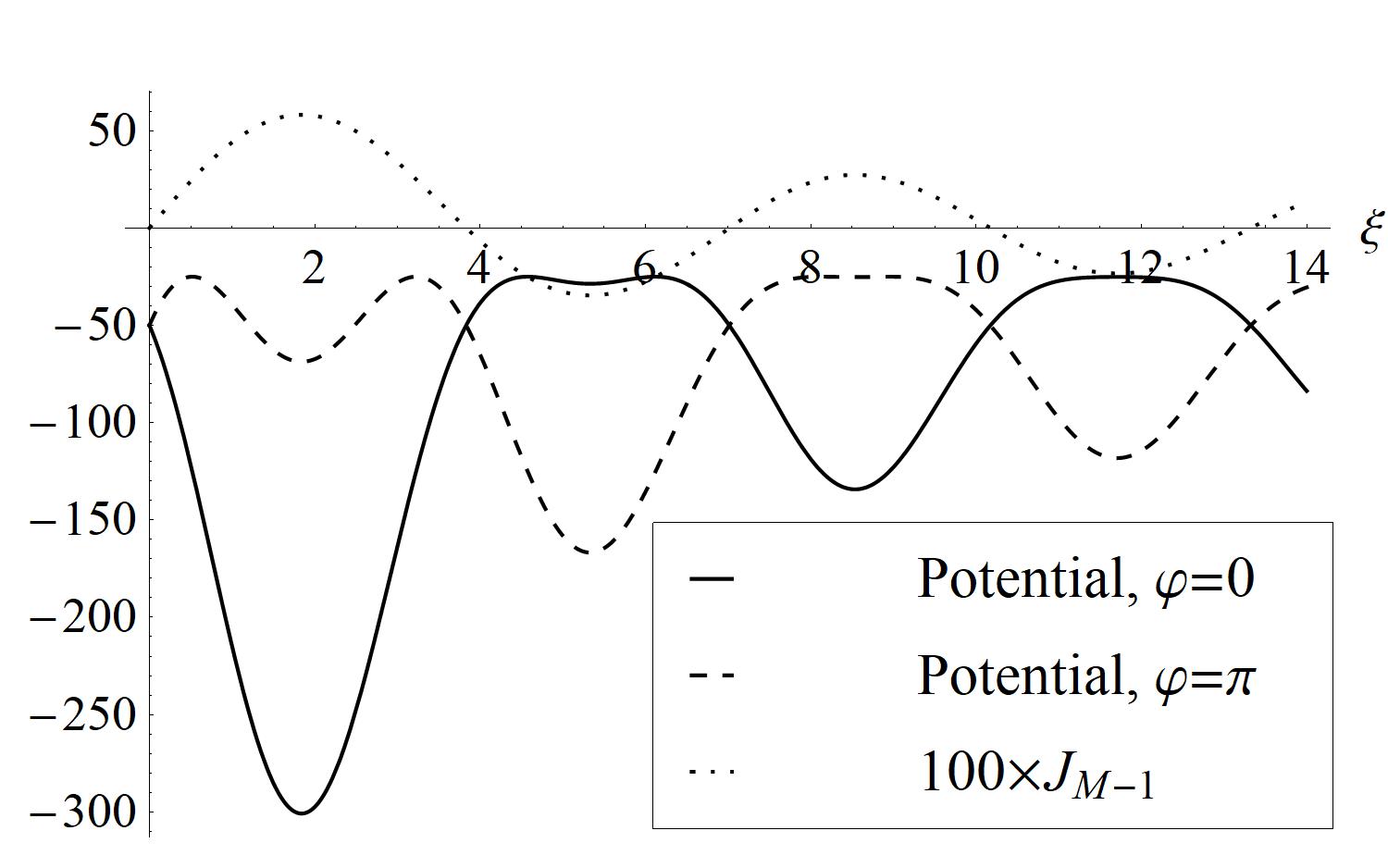}
\end{center}
\caption{Radial dependence of the Bessel function $J_{M-1}$ and of the potential ${\cal V}_a$ in the particular case of $M=2$ for $\zeta=0$, $q=10$ and $\varphi=0$ (solid line) or $\varphi=\pi$ (dashed line). It is visible that for $\varphi=0$ the minima of the potential correspond to the maxima of the Bessel function and  to the minima of the Bessel function for $\varphi=\pi$. In turn the maxima of the potential are localized close to the minima of the Bessel function for $\varphi=0$ and conversely for $\varphi=\pi$. The only exception appears close to the first maximum of the Bessel function but it is getting shallower as $q$ is growing and finally disappears for $q>20$. For greater clarity, the Bessel function values are scaled $100$ times.}
\label{potebe}
\end{figure}

When the circular polarization of the plane wave is negative with respect to the direction of the beam then the large term with $(1+\chi)$ disappears and ${\cal V}_{as}$ takes the form
\begin{eqnarray}
&&\!\!\!\!\!\!\!{\cal V}_{as}({\bm \xi},\zeta)=-\kappa_+^2J^2_{M-1}(\xi)\label{vcaltsab}\\
&&\!\!\!\!-2q\kappa_-J_{M+1}(\xi)\cos\left(\frac{1-\sigma_z}{\sigma_z}\,\zeta-(M+1)\varphi+\psi\right).
\nonumber
\end{eqnarray}
However, in order to justify the approximation applied one should have $q\kappa_-\gg 1$, which implies $q\gg 20$ for the data used in the present work. If this is not satisfied, only the first term survives. From the practical point of view this case is, then, uninteresting since it virtually gives the same results as the potential~(\ref{vcals}).  

For a fixed $\zeta$ (for instance for $\zeta=0$) the extremes of ${\cal V}_{as}$ can be found by requiring  $\partial{\cal V}_{as}/\partial\xi=0$ and $\partial{\cal V}_{as}/\partial\varphi=0$. The more detailed analysis completed in a way very similar to that given in Sec.~\ref{trl} (but with the role of variables $\varphi$ and $\zeta$ interchanged and for arbitrary $M$) leads to the following general conclusions: 

\begin{enumerate}
\item Deep minima of ${\cal V}_{as}$ are located at maxima of the Bessel function $J_{M-1}(\xi)$ and for $(M-1)\varphi+\psi=2n\pi$ (for integer $n$) and also at minima of the Bessel function and $(M-1)\varphi+\psi=(2n+1)\pi$. Higher values of $\varphi$ (exceeding $2\pi$) must be allowed due to the $\zeta$ dependence in (\ref{vcaltsa}). 
\item The strong maxima of ${\cal V}_{as}$ are located at minima of the Bessel function and for $(M-1)\varphi+\psi=2n\pi$ or at maxima of the Bessel function and $(M-1)\varphi+\psi=(2n+1)\pi$. From the practical point of view, they correspond to low energy concentrations and can serve to trap atoms in the blue-detuned case. 
\end{enumerate}

\begin{figure}[h]
\begin{center}
\includegraphics[width=0.475\textwidth,angle=0]{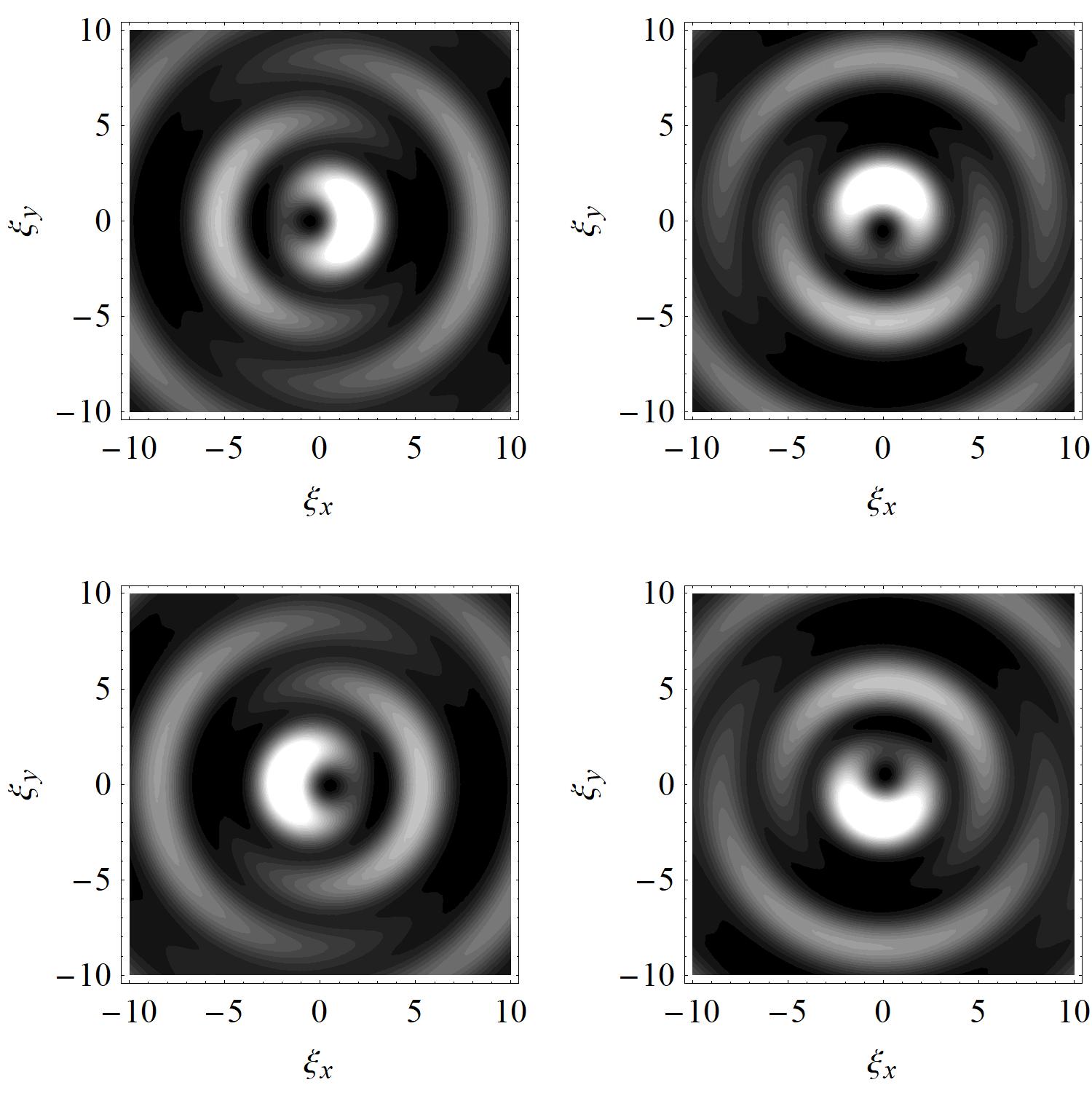}
\end{center}
\caption{The surface graphics representing the cuts of the function ${\cal V}_{as}({\bm \xi}, \zeta)$ with the planes: $\zeta=0, 300, 600, 900$, for $M=2$ and $q=10$ (or $q=5$ in the case of the circular polarization.)}
\label{rotation}
\end{figure}

For considered values of $q$ the additional shallow minimum appears at first maximum of the Bessel function, but this is inessential for the conclusions of our work. All observations can be confirmed when looking at Fig.~\ref{potebe}.

\begin{figure}[h]
\begin{center}
\includegraphics[width=0.3842\textwidth,angle=0]{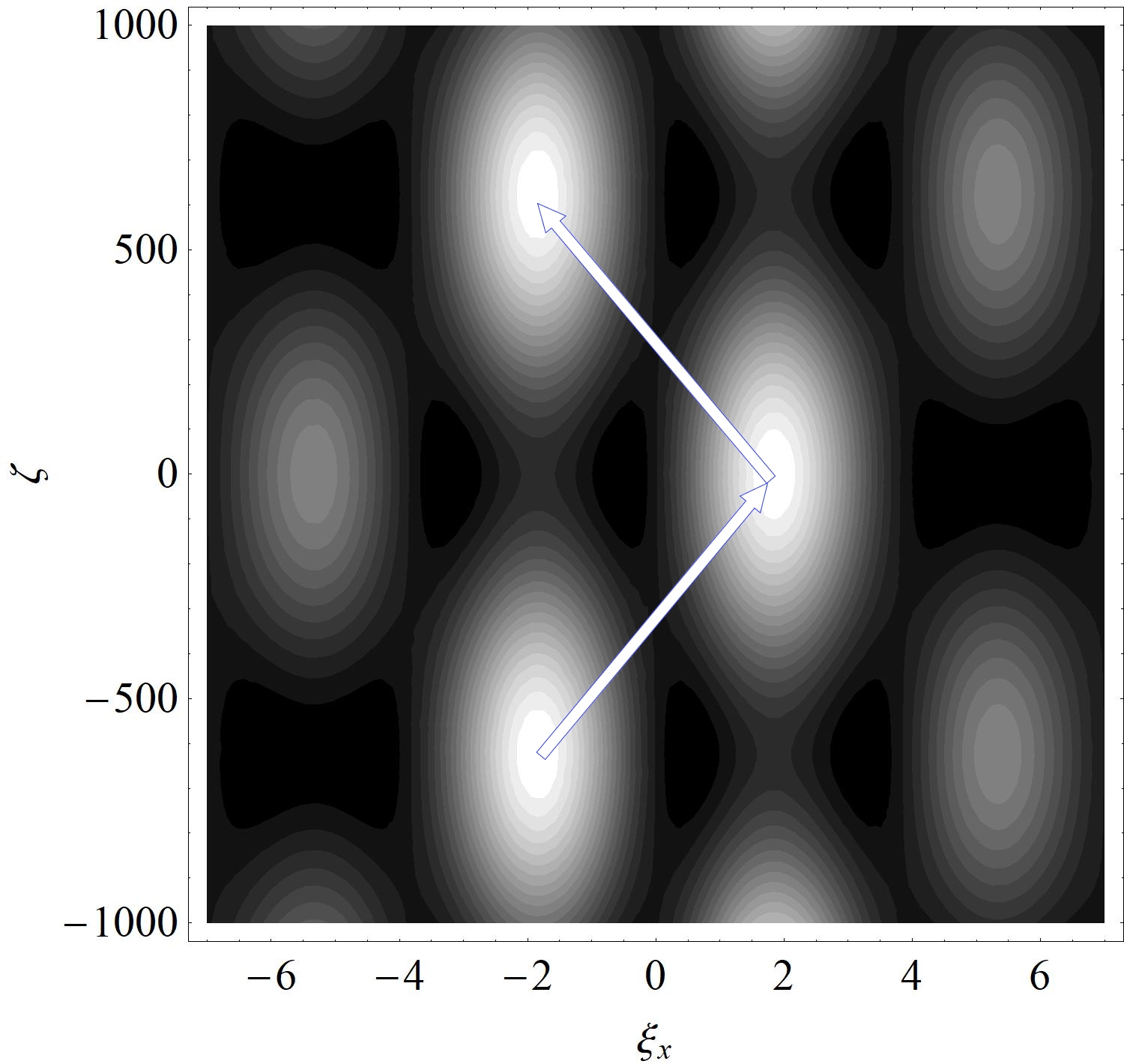}
\end{center}
\caption{Cut of the graph of the function ${\cal V}_{as}({\bm \xi}, \zeta)$ with the plane $\xi_y=0$. Other parameters are identical as in Fig.~\ref{rotation}. The arrows show the path of the potential valley.}
\label{pion}
\end{figure}

In order to reproduce the spatial character of the potential valleys, the diagrams shown in Fig.~\ref{rotation} are prepared. The phase $\psi$ which, to some extent, can serve to manipulate guided atoms but is not an important issue from the point of view of the present work, is henceforth set to zero. The diagrams represent intersections of the function ${\cal V}_{as}({\bm \xi}, \zeta)$ for $M=2$ with horizontal planes for sequentially increasing values of $\zeta$. As earlier, bright regions correspond to high energy concentrations, i.e., the potential dimples (from the point of view of atoms in a red-detuned wave). It is visible from the grayscale level that subsequent minima are shallower, which stays in agreement with the plot of Fig.~\ref{potebe}. On the first diagram all the brightenings are located on the $\xi_x$-axis in line with our former analysis. On the subsequent diagrams when raising the horizontal plane the rotation of the potential minima is observed. Since in fact $\zeta$ is a continuous variable, instead of isolated points these have the form the entire helices.

Similar conclusions can be formulated in relation to dark regions corresponding to the low energy density connected with the maxima of ${\cal V}_{as}$. They are also arranged in the form of helices which allow atoms to be guided when blue-detuned. 

Fig.~\ref{pion} shows the cut of the graph of ${\cal V}_{as}({\bm \xi}, \zeta)$ made with the vertical plane ($\xi_y=0$). When rotating the diagram around the $\zeta$-axis according to the right-hand rule, all bright regions are moving up (as marked with arrows in the figure). The situation would be opposite if the Bessel beam carried the negative vorticity (i.e., for $M-1<0$).

The identified potential valleys in the form of helices can serve to guide neutral atoms. For each helix the value of the pitch can be obtained from~(\ref{vcaltsa}) in the form of

\begin{equation}
\mathrm{pitch}=2\pi(M-1)\, \frac{\sigma_z}{1-\sigma_z},
\label{skok}
\end{equation}
i.e., depends exclusively on the parameters characterizing the Bessel beam and not guided particle. 
Since $\sigma_z$ is close to one, the helix pitch is a large number. For instance taking $\sigma_\perp=0.1$ one gets
\begin{eqnarray}
\sigma_z&\!\!\!=&\!\!\!\sqrt{1-\sigma_\perp^2}=\sqrt{0.99}\approx 0.995,\label{hp}\\
&&\mathrm{and}\;\; \mathrm{pitch}\approx 1247\times(M-1)\nonumber
\end{eqnarray}
in units of $k_z^{-1}$, which means that the pitch amounts to a few hundreds of micrometers. This constitutes a significant difference with respect to helices obtained in \cite{ks} by combining two Bessel beams. In that case the pitch was given as comparable to the wavelength, i.e., it was very small, which makes it harder to simultaneously guide multiple atoms (as will be discussed below) without considering their mutual interactions. The helices become then extremely tight as can be seen in Fig.6 of \cite{ks}. This feature practically eliminates the possibility of guiding micro-objects.

It should be particularly emphasized that these helices (and consequently the trajectories of particles) are fixed in space. They are not affected by the initial conditions as was that of Fig.~\ref{3db}. They even do not depend on which atoms one is dealing with. Neutral particles are guided along the {\em identical} potential valleys as long as their initial energy is not too great to spoil the entire trap. Naturally the potential depth is atom dependent due to different values of polarizability. 

\begin{figure}[h]
\begin{center}
\includegraphics[width=0.475\textwidth,angle=0]{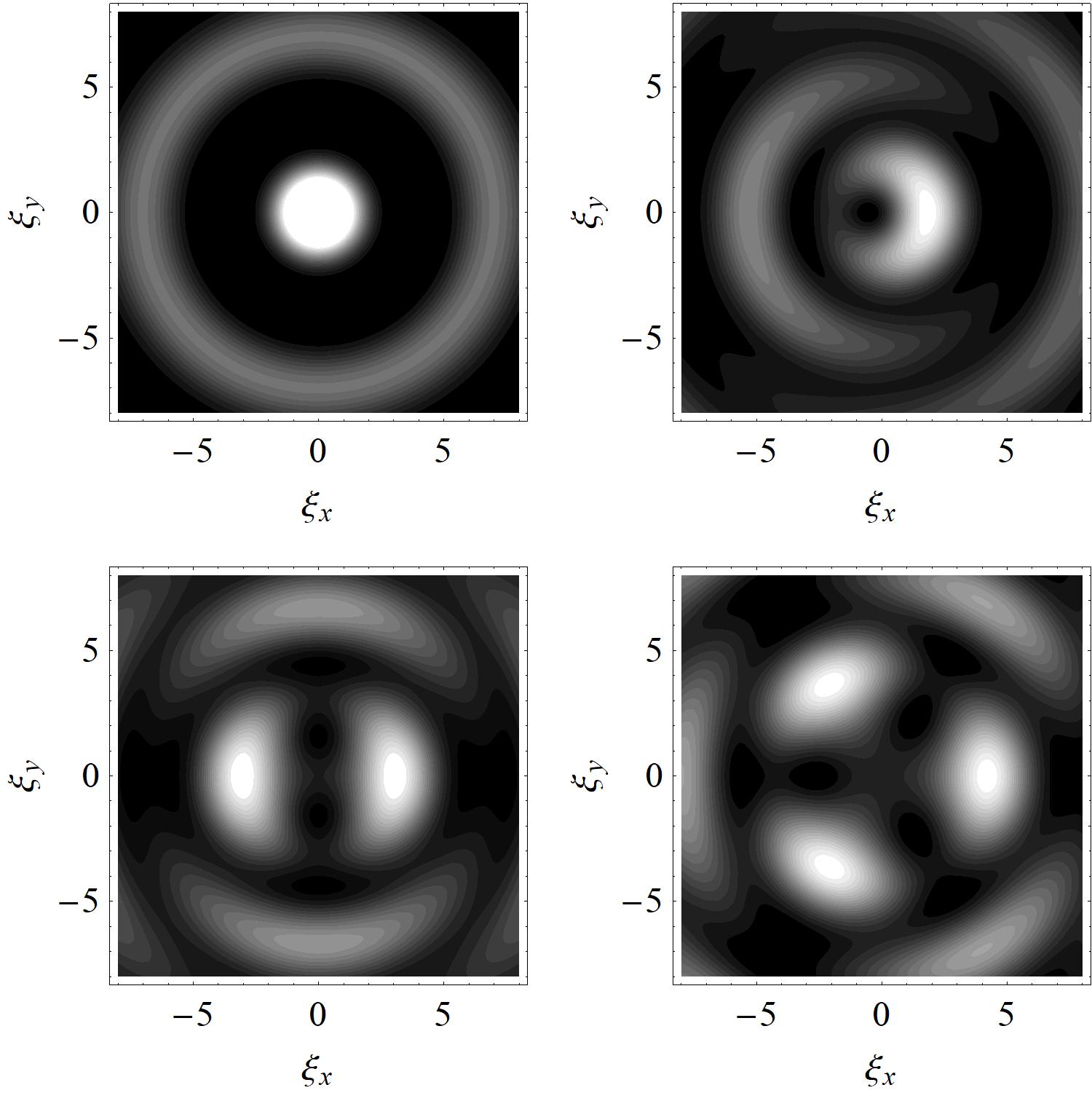}
\end{center}
\caption{The surface graphics representing the cuts of the function ${\cal V}_{as}({\bm \xi}, \zeta)$ with the plane $\zeta=0$ and $q=10$ for the Bessel beams carrying vorticity $0$, $1$, $2$ and $3$ (i.e., $M=1,2,3,4$). It is visible that exactly the same number of potential valleys is generated. They are schematically drawn in Fig.~\ref{tunp}.}
\label{vortices}
\end{figure}

\begin{figure}[h]
\begin{center}
\includegraphics[width=0.5\textwidth,angle=0]{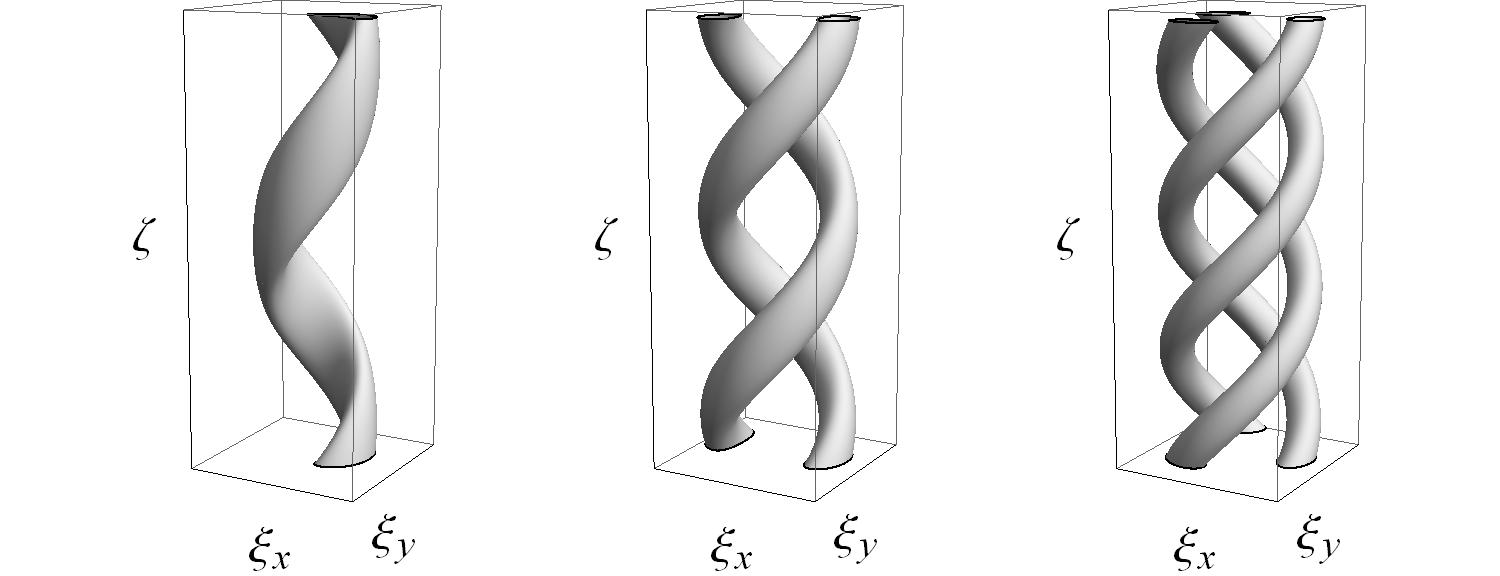}
\end{center}
\caption{Schematic representation of potential valleys for the Bessel beam of vorticity $1$, $2$ and $3$ ($M=2,3,4$).}
\label{tunp}
\end{figure}

The form of the potential valleys shown in Fig.~\ref{rotation} is typical for the Bessel function with vorticity $1$. For such a wave, there is only one bright area on the inner ring and only one helix is built on it. On each subsequent ring there is also only one helix (corresponding, however, to the shallower potential valley). The corresponding diagrams are shown in Fig.~\ref{vortices} and in a schematic way in Fig.~\ref{tunp}. It can be seen that, in general,  on each Bessel ring as many helical tubes, which surround each other, are built as the vorticity of the beam amounts. Each helical tube corresponds to the potential valley and can serve to guide atoms of positive polarizability. Similar tubes, but corresponding to weaker potential wells, are located on the outer rings. They can be shifted along the $\zeta$-axis by modifying the relative phase $\psi$ in~(\ref{elepw}). This property can help to manipulate the trapped objects.

Identical (helical) structure as sketched in Fig.~\ref{tunp} can be drawn for dark regions, i.e. for the potential valleys of the atoms in the blue-detuned beam.

When superposing two Bessel beams, propagating in opposite directions, the whole system of similar helical tubes occurs~\cite{ks} with all the aforementioned differences. If these beams are perpendicular a true $3D$ trap can be obtained (for charged particles)~\cite{ibb1}.

It should be noted that the helical potential tubes are not the exclusive property of Bessel beams. A comparable potential pattern, for a two-level atom, was found for the superposition of two counter-propagating Laguerre-Gaussian beams of opposite helicity~\cite{rsheedb}. Such atoms can then be transported if the entire helical structure is made rotating due to the slight detuning of the beams frequencies~\cite{rsheeda}. The same effect in our case was mentioned at the beginning of this section. Similarly as for the superposition of Bessel beams the emerging helices are very tight. This can be seen in Fig. 1 of \cite{rsheedb}.

\subsection{Helical motion}\label{hm}

When a Bessel beam interferes with a plane wave the `screw' symmetry connected with the simultaneous shift of $\zeta$ and $\varphi$ which led to the constant of motion (\ref{const}) in the pure Bessel case, is lost. There occurs only an approximate constant stemming from the form of the potential~(\ref{vcaltsa}):

\begin{equation}
{\cal C}_{as}=\frac{1}{\beta}\left(\xi_x\dot{\xi}_y-\dot{\xi}_x\xi_y\right)+\frac{1}{\gamma}(M-1)\frac{\sigma_z}{1-\sigma_z}\, \dot{\zeta}.
\label{constas}
\end{equation}

Comparing~(\ref{const}) and~(\ref{constas}) one can notice that the plane wave contributed one unit of AM and presently ${\cal C}_{as}$ reduces to the $\zeta$ component of the particle's AM for $M=1$. This is not strange, since~(\ref{constas}) is derived from {\em the short-time averaged potential} for which, as was discussed previously in the pure Bessel case, this component of particle's AM is conserved (see~(\ref{vcals})). Consequently the only contribution in this limit has to come from the interference between the plane and Bessel waves, which becomes $\varphi$ independent only if the Bessel beam is not carrying any vorticity (i.e., for $M=1$). If the particle accelerates upward for the positive beam vorticity, its AM decreases, if the vorticity is negative - it increases. This transfer of the angular momentum to or from the atoms is the effect that can help in manipulating them.

Apparently, the conservation of particle's AM seems not to depend on the strength of the plane wave, since in ${\cal C}_{as}$ there is no explicit $q$ dependence. This, however, is not true, because the deviation from the uniform motion in the $\zeta$ direction is related to the value of $q$. For weak plane-wave field $\dot{\zeta}$ is almost a constant (in the above-discussed average sense) leading to the similar conclusion as in the pure Bessel case. For stronger field the interference effects become more relevant.

\begin{figure}[h]
\begin{center}
\includegraphics[width=0.475\textwidth,angle=0]{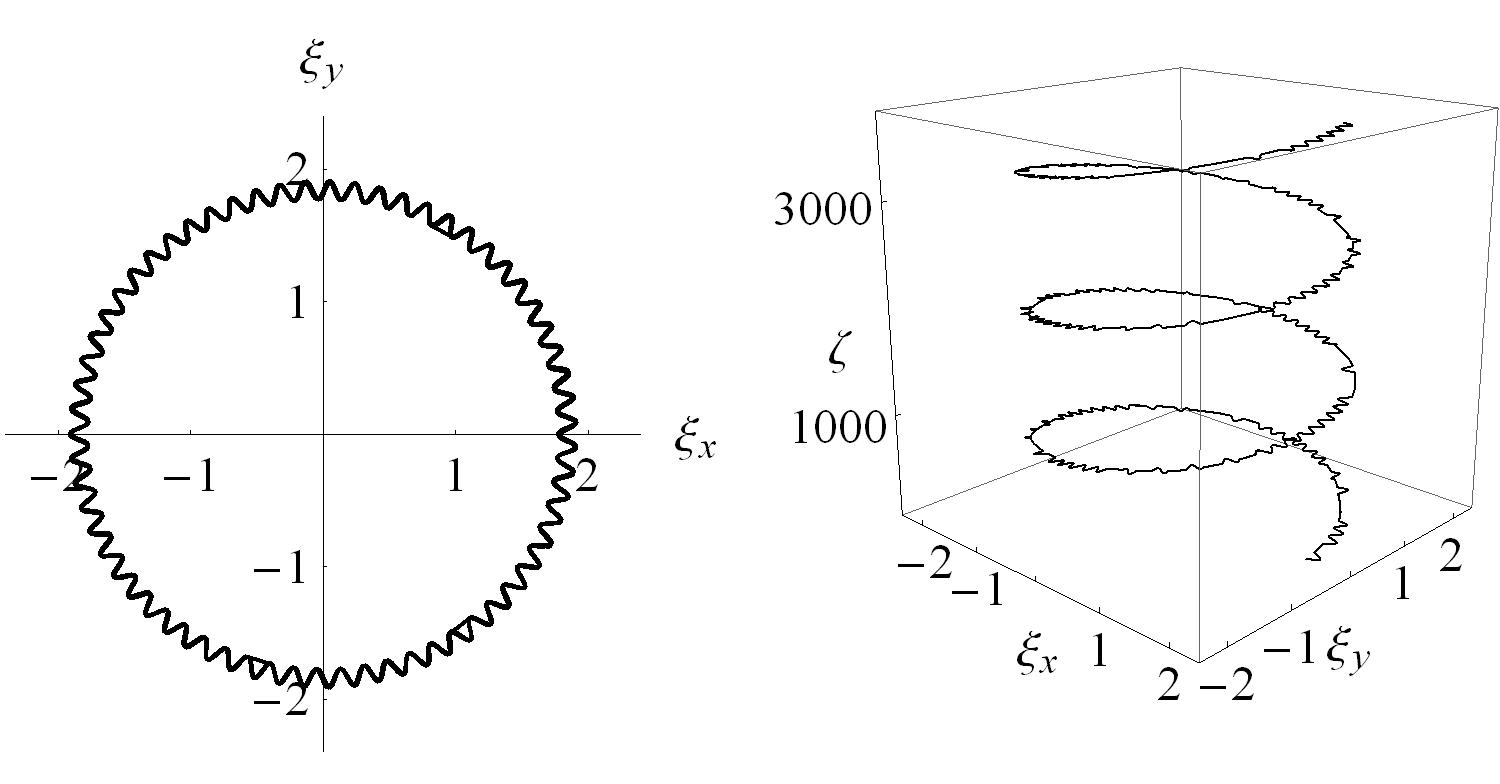}
\end{center}
\caption{The trajectory of an atom in the red-detuned Bessel beam of $M=2$ interfering with the linearly polarized plane wave. Here $\beta = 10^{-5}$, $\sigma_\perp=0.1$, $q=10$.}
\label{heli1}
\end{figure}

In Fig.~\ref{heli1} the trajectory of an atom in the red-detuned Bessel beam of $M=2$ interfering with a plane wave is drawn. It is obtained by means of the numerical integration of equations~(\ref{rrdiml}) with the full potential~(\ref{vcalt}) on the right-hand sides. Apart from the local deviations the atom is forced to follow the helical potential valley shown in the first picture of Fig.~\ref{tunp}. As already told before, contrary to the helix of Fig.~\ref{3db} whose radius was set by the extension of the Bessel ring but the motion along the beam was entirely unrestricted and depended on the initial state of the particle, the helix of Fig.~\ref{heli1} is fixed by the parameters of interfering waves only. For instance the pitch of any trajectory exactly equals to the value~(\ref{hp}). 

\begin{figure}[h]
\begin{center}
\includegraphics[width=0.475\textwidth,angle=0]{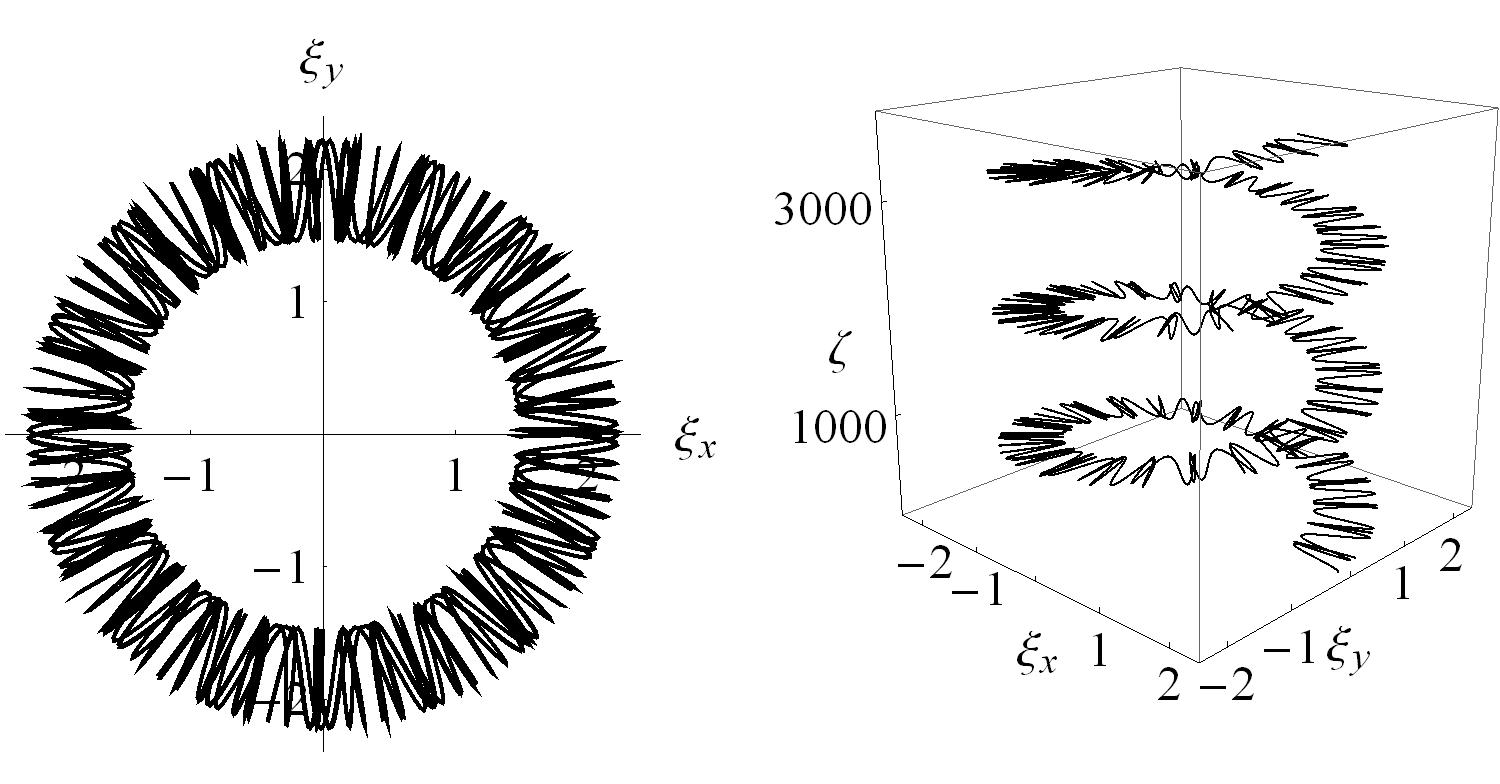}
\end{center}
\caption{Same as Fig.~\ref{heli1}, but for different initial conditions for the atom.}
\label{heli2}
\end{figure}

It would be interesting to see how this trajectory is affected by the initial data of the atom. Comparing the  results of numerical calculations presented in Fig.~\ref{heli1} and in Fig.~\ref{heli2} which have been obtained for different initial states but for the identical combination of fields, it can be easily observed that the modifications only apply to the local motion of the atom. Microscopically the atom moves along a different trajectory but in both cases in a visible way the particles are guided strictly along the same helical valleys of the potential determined by the approximate equation~(\ref{vcaltsa}). This phenomenon is of practical significance for maintaining atoms in stable trajectories. Naturally, if the initial energy were too large, this property would be nullified and the particle would escape from the trap. The depth of this potential well can be estimated. For instance for the laser beam of intensity $I=10^{11}\,\mathrm{W/cm^2}$ and for $q=10$ (or $q=5$ in the case of circular polarization) it amounts to about $8.4\,\mathrm{eV}$. 

\begin{figure}[h]
\begin{center}
\includegraphics[width=0.475\textwidth,angle=0]{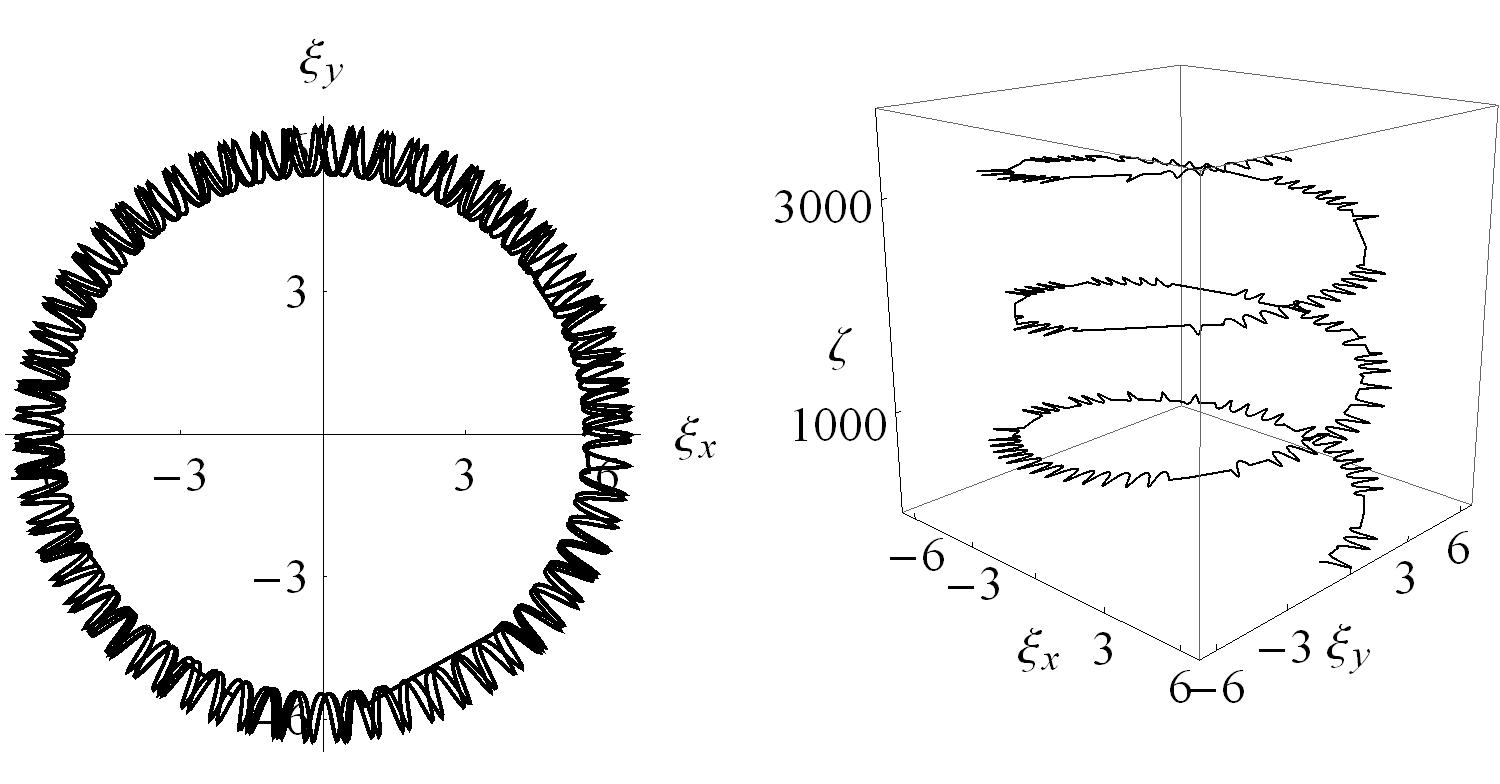}
\end{center}
\caption{The trajectory of an atom in the blue-detuned Bessel beam of $M=2$ interfering with the linearly polarized plane wave. Here $\beta = -10^{-5}$, $\sigma_\perp=0.1$, $q=10$.}
\label{heli3}
\end{figure}

Maxima of the potential ${\cal V}_{as}$, i.e. minima of the energy density (dark regions shown in Figures~\ref{rotation} and~\ref{vortices}), can serve as traps for atoms in blue-detuned fields. In the case of charged particles they correspond to the minima of the ponderomotive potential~\cite{ibb1}. The system of potential valleys is, in this case, quite analogous to that shown in Fig.~\ref{tunp}. The exemplary trajectory of an atom in the potential ${\cal V}$ for $\beta<0$, i.e. for the blue-detuned Bessel beam (of vorticity $1$) is presented in Fig.~\ref{heli3}. As one can see, atoms follow similar trajectories, but the depth of the potential well can be estimated to be about a half of what was obtained for red-detuned beam.  

\begin{figure}[h]
\begin{center}
\includegraphics[width=0.28\textwidth,angle=0]{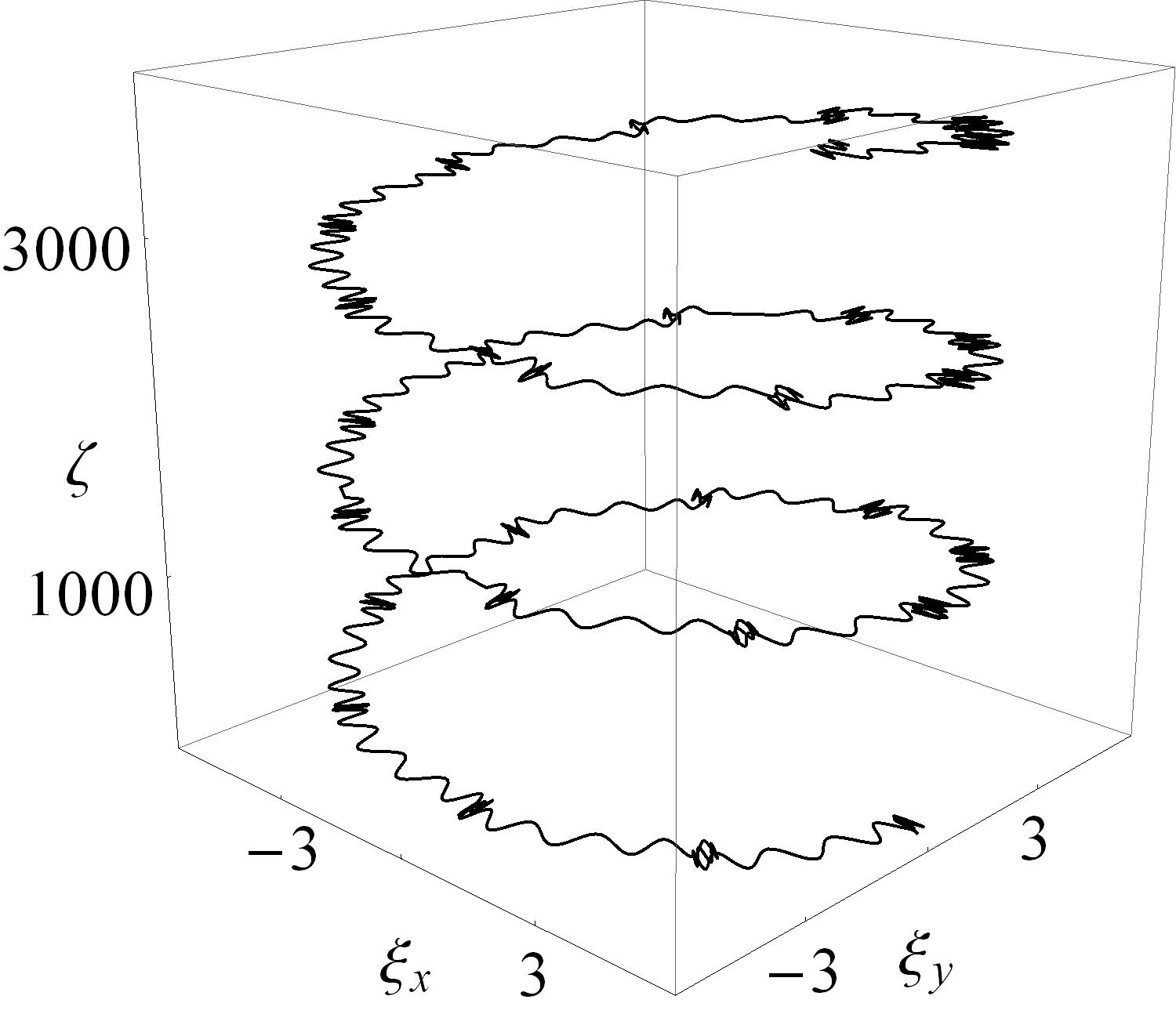}
\end{center}
\caption{The trajectory of an atom with the same initial data as in Fig.~\ref{heli1} but placed in the Bessel beam of vorticity $-1$ ($M=0$) interfering with a plane wave.}
\label{heli4}
\end{figure}

The opposite twist of helices, if needed, is not achieved in this setup by modifying the initial state of particles (as would be in the case described in Sec.~\ref{bb}) on which we have little control. To this goal one should rather use a Bessel beam with negative vorticity. Fig.~\ref{heli4} shows the trajectory of an atom with the same initial state as that of Fig.~\ref{heli1} (obviously, its initial position must fit to the new potential valley) in the case of the Bessel beam with vorticity equal to $-1$. Note, that the atom is forced to follow a new helix with the opposite twist created by the new combination of electromagnetic fields. This provides us with the ability to steer the motion of trapped particles.

\begin{figure}[h]
\begin{center}
\includegraphics[width=0.48\textwidth,angle=0]{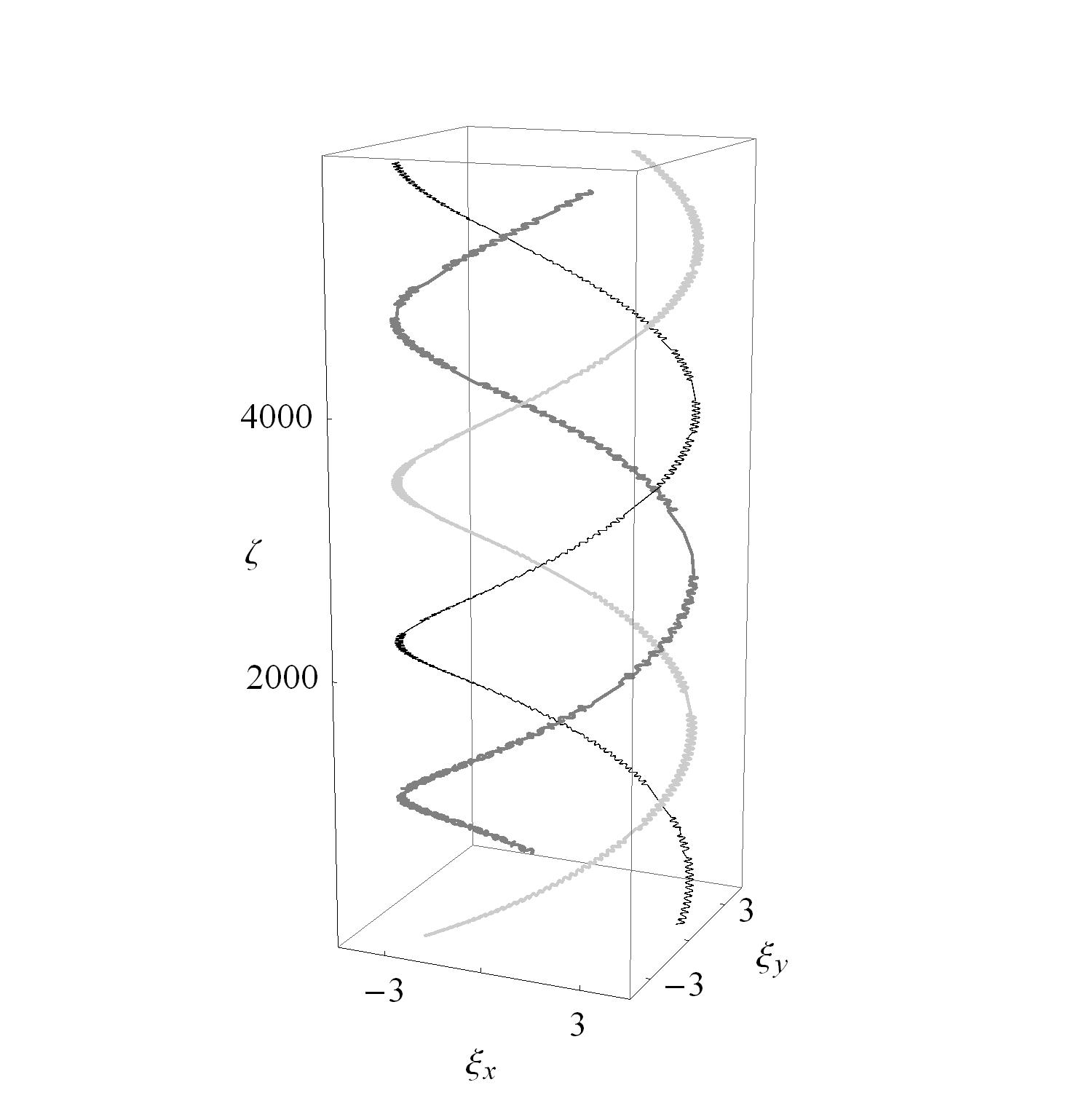}
\end{center}
\caption{Three particles performing independent motion along three helices created by the Bessel beam of vorticity $3$ ($M=4$) interfering with a plane wave.}
\label{3part}
\end{figure}

The system of potential minima as shown in Figures~\ref{vortices} and~\ref{tunp} allows multiple atoms to be conducted simultaneously, each along its own potential valley. In the case of neutral atoms the interaction between one another is much weaker than that of charged particles moving in the ponderomotive potential and seems to be a real practical possibility. Fig.~\ref{3part} shows the trajectories of three atoms obtained by numerically solving equations in the potential with $M=4$. They precisely reflect the structure of the potential tubes for a Bessel beam with vorticity $3$.

\subsection{$3D$ traps}\label{trl}

\begin{figure*}[t]
\begin{center}
\includegraphics[width=0.95\textwidth,angle=0]{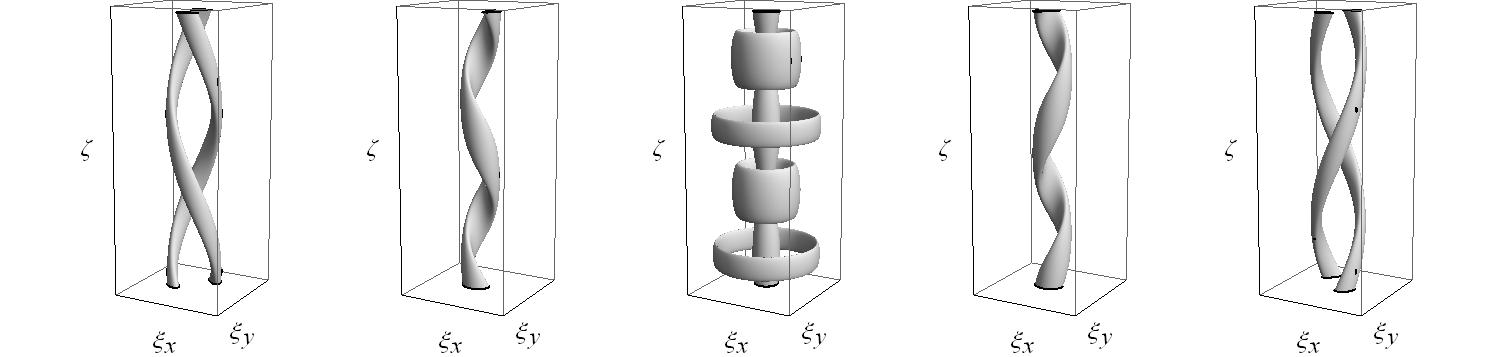}
\end{center}
\caption{Schematic representation of the potential valleys for a Bessel beam of vorticity $-2,-1,0,1,2$ interfering with a plane wave.}
\label{helices}
\end{figure*}

According to the expression~(\ref{skok}) the pitch of the helical potential valleys is proportional to $M-1$. For smaller values helices are more tight and finally become entirely degenerated if $M=1$ (no vorticity). Instead of extended helices, the potential wells have now the character of true, three-dimensional rings (and not of cylindrical structures as in~Sec.~\ref{bb}). They are depicted in Fig.~\ref{helices}. Each of these visible rings can serve as a real $3D$ trap for neutral atoms in the red-detuned case. Similar toroidal structures can be obtained by interfering two Bessel beams~\cite{ks}.

The spatial position of these ring-shaped minima can be found by fixing $\varphi$ and solving the set of equations ($\phi$ is set to zero for simplicity):
\begin{subequations}\label{dec}
\begin{align}
&\frac{\partial {\cal V}_{as}}{\partial \xi}= -\kappa_+J_0'(\xi)\left[2\kappa_+J_0(\xi)+q\cos\left(\frac{1-\sigma_z}{\sigma_z}\,\zeta\right)\right]=0,\label{decxi}\\
&\frac{\partial {\cal V}_{as}}{\partial \zeta}= q\kappa_+J_0(\xi)\frac{1-\sigma_z}{\sigma_z}\,\sin\left(\frac{1-\sigma_z}{\sigma_z}\,\zeta\right)=0.\label{deczeta}
\end{align}
\end{subequations}
Leaving aside the manifestly contradictory case, it is clear that the following obvious possibilities occur:
\begin{subequations}\label{pos}
\begin{align}
&J_0(\xi)=0\;\;\;\mathrm{and}\;\;\; \cos\left(\frac{1-\sigma_z}{\sigma_z}\,\zeta\right)=0,\;\;\;\;\mathrm{or}\label{pos1}\\
&J_0'(\xi)=0\;\;\;\mathrm{and}\;\;\; \sin\left(\frac{1-\sigma_z}{\sigma_z}\,\zeta\right)=0,\;\;\;\;\mathrm{or}\label{pos2}\\
&\sin\left(\frac{1-\sigma_z}{\sigma_z}\,\zeta\right)=0\;\;\;\mathrm{and}\label{pos3}\\ 
&\hspace*{10ex} J_0(\xi)=-\frac{q}{2\kappa_+}\,\cos\left(\frac{1-\sigma_z}{\sigma_z}\,\zeta\right)\nonumber.
\end{align}
\end{subequations}

Analyzing these equations together with the matrix of second derivatives of ${\cal V}_{as}$, it can be shown that there emerge two kinds of the ring-shaped potential minima (both numbered with an integer $n$):
\begin{enumerate}
\item Those deeper located at 
\begin{equation}
\zeta=(2n+1)\pi(\sigma_z)/(1-\sigma_z)
\label{}
\end{equation}
with the depth
\begin{equation}
{\cal V}_{\mathrm{min}}^{(1)}=-\kappa_+J_0(\xi_\mathrm{min})\left[\kappa_+J_0(\xi_\mathrm{min})-q\right],
\label{vmin1}
\end{equation}
where $\xi_\mathrm{min}\approx 3.832$ stands for the position of the first minimum of the Bessel function $J_0$, and
\item those shallower located at $\zeta=2n\pi(\sigma_z)/(1-\sigma_z)$ with the depth
\begin{equation}
{\cal V}_{\mathrm{min}}^{(2)}=-\kappa_+J_0(\xi_\mathrm{max})\left[\kappa_+J_0(\xi_\mathrm{max})+q\right],
\label{vmin2}
\end{equation}
where $\xi_\mathrm{max}\approx 7.016$ is the position of the first (nonzero) maximum of the Bessel function $J_0$.
\end{enumerate}

\begin{figure}[h]
\begin{center}
\includegraphics[width=0.475\textwidth,angle=0]{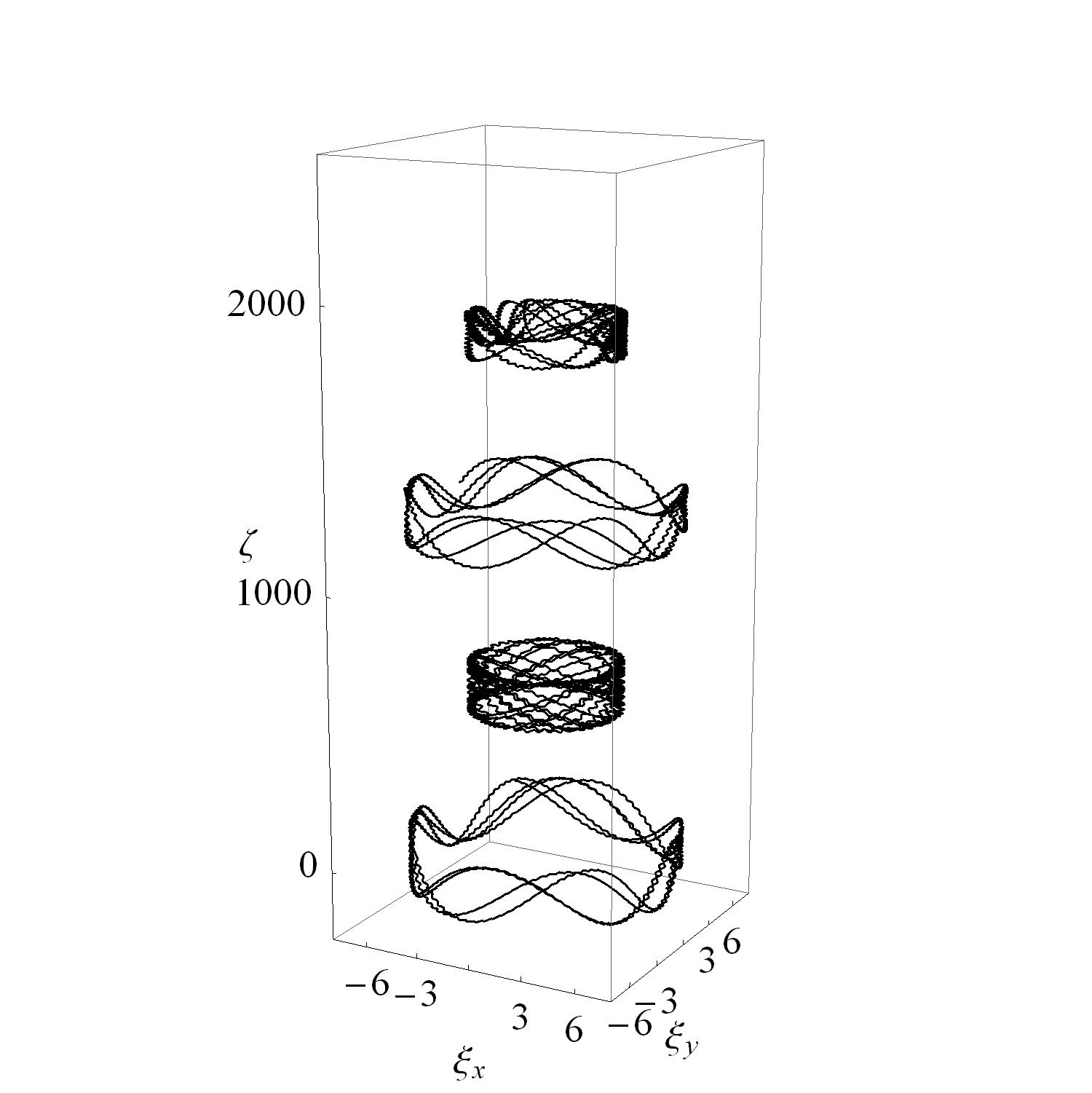}
\end{center}
\caption{Trajectories of four atoms independently trapped in four ring-shaped potential valleys of the third diagram in Fig.~\ref{helices} (i.e., for $M=1$). The values of parameters are identical as before: $\beta = 10^{-5}$, $\sigma_\perp=0.1$, $q=10$. }
\label{3Dtrap}
\end{figure}

For the data used throughout the paper ($q=10,\sigma_\perp=0.1$) it can be found that
\begin{subequations}\label{val}
\begin{align}
&{\cal V}_{\mathrm{min}}^{(1)}\approx 144.9\;\;\;\;\mathrm{at}\;\;\;\; \zeta=623.6,\, 1870.8,\;\ldots\label{v1}\\
&{\cal V}_{\mathrm{min}}^{(2)}\approx 95.7\;\;\;\;\mathrm{at}\;\;\;\; \zeta=0,\, 1247.2,\;\ldots\label{v2}
\end{align}
\end{subequations}
which entirely agrees with Figures~\ref{helices} and~\ref{3Dtrap}. Assuming as before the intensity of the beam to be $I=10^{11}\,\mathrm{W/cm^2}$ one gets that the depths of these minima can be estimated to $5.22\,\mathrm{eV}$ and  $3.44\,\mathrm{eV}$ correspondingly.

Fig.~\ref{3Dtrap} shows the trajectories of several atoms simultaneously trapped in ring-shaped potential wells. The obtained trajectories in the visible way correspond to the potential rings of the third diagram in Fig.~\ref{helices}.

The mutual interaction of dipole moments of polarized neutral atoms trapped in different rings are negligible, being many orders of magnitude weaker than the interaction with electromagnetic fields. The spatial separation of subsequent ring-shaped traps is of order of $\Delta z\approx \Delta \zeta\, c\,\omega^{-1}$, which amounts to hundreds of micrometers for $\omega\lesssim 10^{15}\,\mathrm{Hz}$. As compared to the rings obtained in \cite{ks}, they are thousands times further apart from each other (see Eq. (11) in the quoted work). 
Therefore, in the presently proposed fields configuration, one can truly consider the parallel guiding or trapping atoms in their individual $2D$ or $3D$ traps. This situation would not be so comfortable for charged particles in a ponderomotive potential, where their interaction has to be taken into account.

\section{Summary}\label{sum}

The present work is concerned with the motion of neutral polarizable particles (atoms) in the field of a pure Bessel beam and in the combined fields of a Bessel beam and a plane wave propagating in the same direction. The interaction potential is proportional to ${\bm E}^2$. This causes the particles with positive polarizability (red-detuning) to be dragged into the areas of high energy density, and those with negative polarizability (blue-detuning) to be pushed out of them. It is known that in the Bessel beam alone there arise concentric rings in the plane perpendicular to the propagation axis, where the density of electromagnetic energy is raised, and between them rings with lower energy density are formed. The trajectories of particles obtained from the equations of motion by the numerical integration show, that atoms are captured in the former when red-detuning occurs and otherwise in the latter ones. Each of the rings can serve as a kind of a $2D$ trap, although the outer rings offer the shallower potential wells. The motion in the direction of the wave propagation is unrestricted, within a good approximation, which allows for the guidance of particles along the beam captured at (almost) fixed distance from the beam core.

The superposition of a Bessel beam and a plane wave of the precisely tuned frequencies leads to other interesting structures that can serve as traps as well. There emerge potential valleys in the form of helices inside which particles are guided independently of their initial energy or velocity within reasonable limits (i.e., kinetic energies not exceeding few electronovolts, naturally depending on the laser intensity). Each Bessel ring develops as many helices as the topological vortex index amounts. The pitch of a given helix and the direction of its twist (and simultaneously of atom's trajectory), is also fixed by the by the value of this index. By changing the vorticity into an opposite one, the same particle can be forced to follow either right-twisted or left-twisted path. These properties can serve to manipulate guided atoms in space. The helices depend on the relative phase of both fields, so this parameter can serve as a separate tool for the precise guidance of atoms as well.

In the special case of a Bessel beam carrying no vorticity all the helices become degenerate. Instead of them there arises a series of true $3D$ traps in the form of rings perpendicular to the direction of propagation of both waves. The trajectories of atoms in these traps have been obtained in the numerical way. 

The rich structure of potential valleys creates the possibility to simultaneously guide atoms along the multi-helix system or to capture them in different ring traps, forming a kind of an optical lattice. The interaction between neutral polarized atoms is marginal in this case. A similar framework has been observed for charged particles (for instance electrons) subject to ponderomotive force stemming from the analogous combination of fields or in the case of neutral particles in the potential generated by two counter-propagating Bessel beams. However, their mutual interactions are not negligible either due to electric forces or extremely tight form of helices. 

In conclusion it should be pointed out that superposing a Bessel beam with other waves opens the possibility to manipulate neutral particles thanks to the polarization mechanism. Several parameters which are at our disposal (vorticity topological number, field strength, relative phase, $z$-component of the wave vector etc.) allow to modify particle's trajectories, within a certain class. One can also think of slight detuning of both waves or the relative inclination of their wave vectors. For larger objects the incorporation of the Brownian motion led to hopping between Bessel rings~\cite{brown}. On the quantum level the tunneling effect should also be estimated although seems negligible in the proposed setup.

\end{document}